\documentclass[fleqn]{2023SCGE}
\usepackage{soul}
\setstcolor{red}
\usepackage{subfigure}
\usepackage{multirow,multicol}

\newcommand\Ek{\mbox{\textit{Ek}}}
\newcommand\Prt{\mbox{\textit{Pr}}}
\newcommand\Ra{\mbox{\textit{Ra}}}
\newcommand\St{\mbox{\textit{St}}}

\newcommand{\widebar}[1]{\mathop{\overline{#1}}\nolimits}

\begin{document}

\ensubject{subject}

\ArticleType{Article}
\SpecialTopic{SPECIAL TOPIC: }
\Year{2023}
\Month{January}
\Vol{66}
\No{1}
\DOI{??}
\ArtNo{000000}
\ReceiveDate{January 11, 2023}
\AcceptDate{April 6, 2023}

\title{Retarded Stellar Dynamo in Tidally Deformed M Dwarfs}{Retarded Stellar Dynamo in Tidally Deformed M Dwarfs}


\author[1,2,7\textdagger]{Song Wang}{{songw@bao.ac.cn}}

\author[3,7{\textdagger}]{Wenbo Li}{}

\author[1]{Henggeng Han}{}

\author[3,4]{Dali Kong}{{dkong@shao.ac.cn}}

\author[1,2,4,5]{Jifeng Liu}{{jfliu@bao.ac.cn}}

\author[6]{Xinlin Zhao}{}

\AuthorMark{Wang S}

\AuthorCitation{Wang S, et al}


\address[1]{Key Laboratory of Optical Astronomy, National Astronomical Observatories, Chinese Academy of Sciences, Beijing 100101, China;}
\address[2]{Institute for Frontiers in Astronomy and Astrophysics, Beijing Normal University, Beijing, 102206, China;}
\address[3]{Shanghai Astronomical Observatory, Chinese Academy of Sciences, Shanghai 200030, China;}
\address[4]{School of Astronomy and Space Science, University of Chinese Academy of Sciences, Beijing 100049, China;}
\address[5]{New Cornerstone Science Laboratory, National Astronomical Observatories, Chinese Academy of Sciences, Beijing, 100012, China;}
\address[6]{Department of Physics, Chongqing University, Chongqing 400044, China}

\address[\textdagger]{Equally contributed to this work}


\abstract{Current studies of stellar dynamos primarily focus on spherical stars, leaving their behavior in distorted stars largely unexplored. 
We utilize stars of varying distortions to examine the relation between stellar cycle periods ($P_{\rm cyc}$) and rotational periods ($P_{\rm rot}$), which are closely linked to dynamo processes.
By analyzing a sample of tidally distorted M dwarfs in cataclysmic variables, we identify an anti-correlation between $P_{\rm cyc}$ and $P_{\rm rot}$, in contrast to the lack of such a relation in single M dwarfs. 
This means that stars with greater deformation have longer cycle periods, suggesting variations in dynamo behavior under non-spherical geometries. Our numerical simulations further reveal that, the thermal convection weakens in highly distorted stars, and subsequently, the differential rotation is also reduced. These effects may lengthen the conversion timescale between poloidal and toroidal magnetic fields, potentially explaining the newly discovered $P_{\rm cyc}$--$P_{\rm rot}$ relation in cataclysmic variables.}



\maketitle


\begin{multicols}{2}

\section{Introduction}
\label{sec:intro}

Most studies of dynamo processes, which govern the generation and evolution of magnetic fields, focus on spherically symmetric stars. It is therefore compelling to investigate how dynamo mechanisms operate in non-spherical stars, where the deformation caused by rapid rotation and tidal forces from companion stars may alter the structure of the convection zone and the profiles of differential rotation. 
Stellar activity cycle is thought to be a result of dynamo processes  
\cite{1955ApJ...122..293P, 1995ApJ...438..269B}. Thus, examining the relation 
\Authorfootnote
between cycle periods ($P_{\rm cyc}$) and rotation periods ($P_{\rm rot}$) in both spherical and non-spherical stars offers a promising approach to probing dynamo processes across different stellar geometries.

Pioneering observational and theoretical studies have shown that the $P_{\rm cyc}$--$P_{\rm rot}$ relation serves as a key constraint on dynamo models  \cite{1984ApJ...287..769N,1996ApJ...460..848B,1999ApJ...524..295S,2017Sci...357..185S,2018A&A...616A..72W,2018A&A...616A.160V}. For example, for stars with rotation periods shorter than $\approx$10 days, a weak positive relation between $P_{\rm cyc}$ and $P_{\rm rot}$ is observed for F, G, and K dwarfs in the $P_{\rm cyc}$/$P_{\rm rot}$ vs. 1/$P_{\rm rot}$ diagram on a log--log scale  \cite{1996ApJ...460..848B,2009A&A...501..703O,2016A&A...595A..12S}, but no such relation has been identified for M dwarfs  \cite{2012ARep...56..716S,2016A&A...595A..12S,2018A&A...612A..89S,2019A&A...621A.126D}. 
The different behavior of the relations was explained as that the dynamos in M stars differ from those in F, G, and K stars. For M stars, which are fully or nearly fully convective, their magnetic fields are likely generated by a distributed dynamo, with magnetic cycles potentially governed by turbulent diffusion and meridional circulation  \cite{2019A&A...622A..40K}.

To investigate dynamo behavior in distorted stars, we compiled a sample of M stars, comprising single dwarfs, dwarfs in wide binaries, and those in close binaries, and examined their $P_{\rm cyc}$--$P_{\rm rot}$ relations. For close binaries, we focused on cataclysmic variable (CV) systems,  each consisting of a white dwarf and a tidally distorted late-type dwarf, which offer a clear advantage over normal close binaries, as both the rotation periods and cycle periods can be unambiguously attributed to the dwarf component. 
The paper is organized as follows. In Section~\ref{sec:sample}, we describe the construction of the sample and data analysis, including the compilation of stars with measured cycle and rotation periods from the literature. Section~\ref{sec:relation} presents the relations between $\log(P_{\rm cyc}/P_{\rm rot})$ and $\log(1/P_{\rm rot})$ for M dwarfs.
In Section~\ref{sec:simulation}, we introduce a binary model to investigate the impact of tidal deformation on stellar thermal convection. Finally, Section~\ref{sec:discussion} provides  discussion and conclusions of our findings.

\section{Sample selection and data analysis}
\label{sec:sample}

\begin{figure}[H]
    \center
    \includegraphics[width=0.48\textwidth]{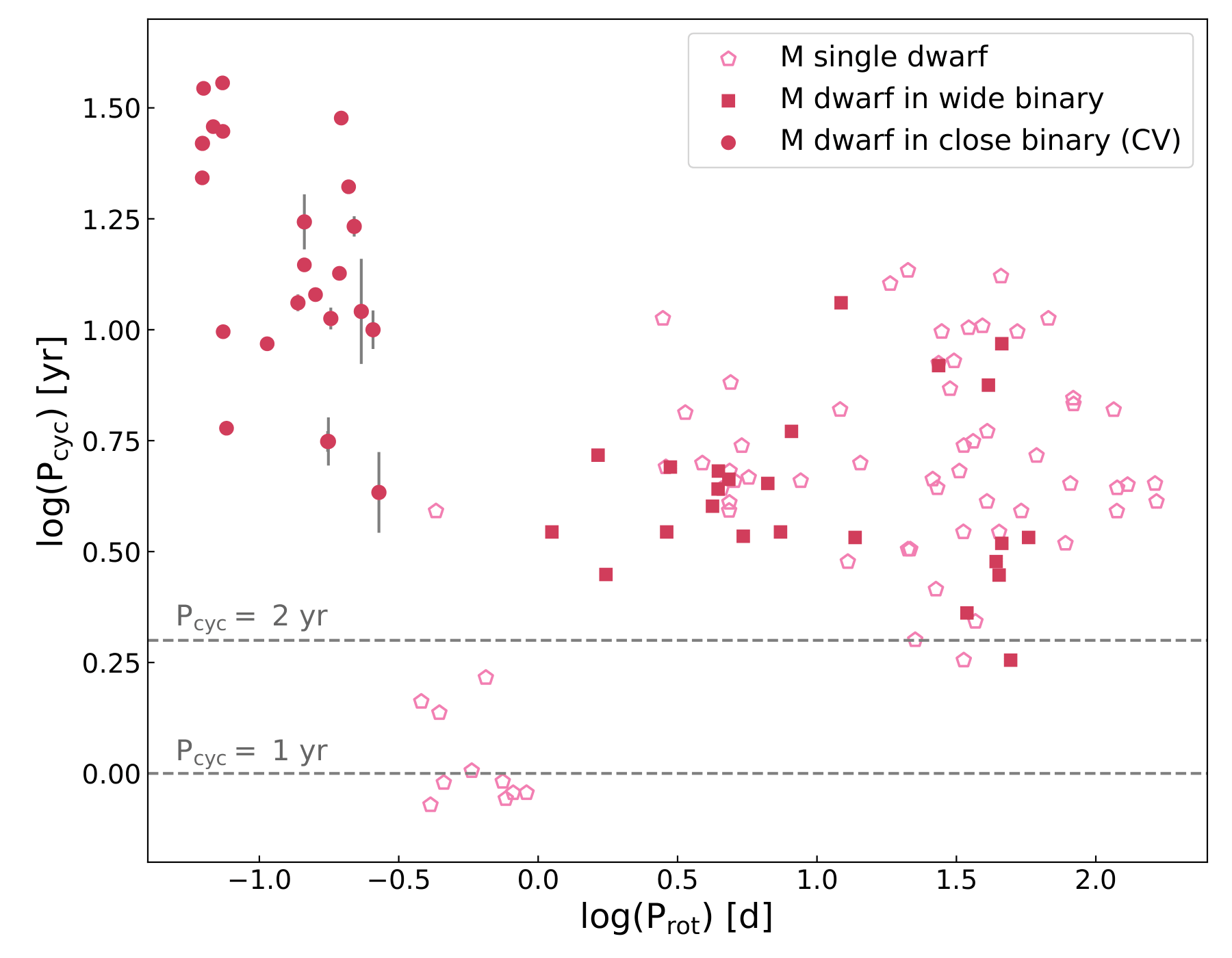}
    \caption{$P_{\rm cyc}$ versus $P_{\rm rot}$ in log--log scale for M dwarfs. Filled dots represent stars in close binaries (i.e., CVs), while filled squares represent stars in wide binaries. Open pentagons represent single stars. The two horizontal dashed lines represent stars with cycle periods of 1 and 2 years, respectively. 
    }
    \label{figure1.fig}
\end{figure}

We compiled a sample of M dwarfs including single stars, wide binaries, and close binaries. 
First, we collected stellar cycle and rotation periods for 89 M singles and M wide binaries from previous studies (e.g., refs. \cite{2013ApJ...764....3R,2013AN....334..972V,2014MNRAS.441.2744V,2016A&A...590A.133O,2016A&A...595A..12S,2018A&A...612A..89S,2016A&A...591A..43D,2017A&A...606A..58D,2019A&A...622A..40K,2019A&A...621A.126D}), which mainly used long-term photometric light curves and chromospheric activity monitoring to identify stellar activity cycles.
Although \cite{2012ARep...56..716S} presented a set of cycle periods for M single stars, those periods have been thought as short secondary cycles and were therefore not included in our sample.
Cross‑matching with SIMBAD identified six M single stars as binaries; they were categorized to the wide binary sample (see Supplementary Material).
This leads to 65 M single stars and 24 M stars in wide binaries. 
Then, we searched for close binaries including M dwarfs, specifically defined as CV systems having orbital periods shorter than 0.3 days.
In these tidally locked systems, the orbital period equals the rotational period of the M dwarf.
The sample is drawn mainly from \cite{1990AJ.....99.1941B,1992ASPC...29..284B,1999ApJ...524..295S,2001A&A...369..882A,2003A&A...404..301R}, yielding 23 M stars in CVs.
The full M-dwarf sample is shown in Figure \ref{figure1.fig}. Detailed information on their rotation and cycle periods is listed in Tables S1--S3.

Generally, in previous studies, standard time-series analysis methods such as (Generalized) Lomb-Scargle periodograms, Fourier transforms, and Phase Dispersion Minimization were employed to detect long-term cyclic variations.
For CVs in our sample, two additional methods were also used to extract cycle periods from light curve data: (1) tracking orbital period variations using the observed-minus-calculated ($O-C$) method \cite{1994PASP..106.1075R}, and (2) measuring variations in the intervals between outbursts in accreting systems \cite{1990AJ.....99.1941B}.
The underlying physical explanation is that magnetic field variations in the convective layers of the cool star (i.e., the M dwarf) during an activity cycle cause the star to expand and contract within a quadrupole-deformed Roche potential \cite{1987ApJ...322L..99A,1988Natur.336..129W}. 
Such radius variations can alter the star's moment of inertia and further produce both detectable apsidal motion and orbital period variations with the same timescale as the magnetic cycle, even in circular orbits \cite{1990AJ.....99.1941B}.
Simultaneously, since the cool star fills its Roche lobe, the changes in radius also result in variation in the mass transfer rate. This, in turn, alters the luminosity of the accretion disk surrounding the white dwarf primary \cite{1990AJ.....99.1941B, 1992ASPC...29..284B}.

We noted that an alternative explanation for the recurring outbursts in CV systems is the disk instability model, which attributes dwarf nova outbursts to thermal and viscous instabilities in the accretion disk, triggered by hydrogen ionization.
However, in old novae and novalike binaries, the accretion disks are typically smaller and more stable than those of dwarf novae. 
More importantly, the approximately regular recurrence times observed in some CVs cannot be explained by the disk instability model. Therefore, in this study, we interpreted the periods measured from variations in outburst intervals as stellar activity cycle periods. 
Fortunately, only four CVs in our sample have their cycle periods determined through the intervals between recurring outbursts.

Similar to our sun, stellar activity cycles can be multiple \cite{2009A&A...501..703O}.
At moderate rotation rates ($\approx$10--20 days), some stars exhibit cycles on two sequences, which were explained as two stellar dynamos operating simultaneously in different shear layers \cite{2007ApJ...657..486B, 2017ApJ...845...79B}.
Different detection methods may also lead to different estimations of cycle lengths.
For example, the cycle lengths unveiled by direct tracking of polarity switches are sometimes significantly shorter than those derived from chromospheric activity monitoring \cite{2011AN....332..866M}.
The cycles identified by CoRoT \cite{2015A&A...583A.134F} or Kepler \cite{2014MNRAS.441.2744V} missions are much shorter ($\lesssim$ 2--3 years) than classical activity cycle lengths.
It should be noted that cycle periods below about 1--2 years, especially those derived from records with approximately yearly gaps, may be highly uncertain \cite{2009A&A...501..703O}.
Therefore, in following analysis, we consider only cycles longer than 2 years, leaving 54 single stars, 23 wide binaries, and 23 CVs (Figure \ref{figure1.fig}).
For stars with multiple reported cycle measurements, we adopted the most recent result to ensure accuracy.

\section{Results: new $P_{\rm cyc}$--$P_{\rm rot}$ relation}
\label{sec:relation}

\begin{figure*}
    \center
    \includegraphics[width=0.98\textwidth]{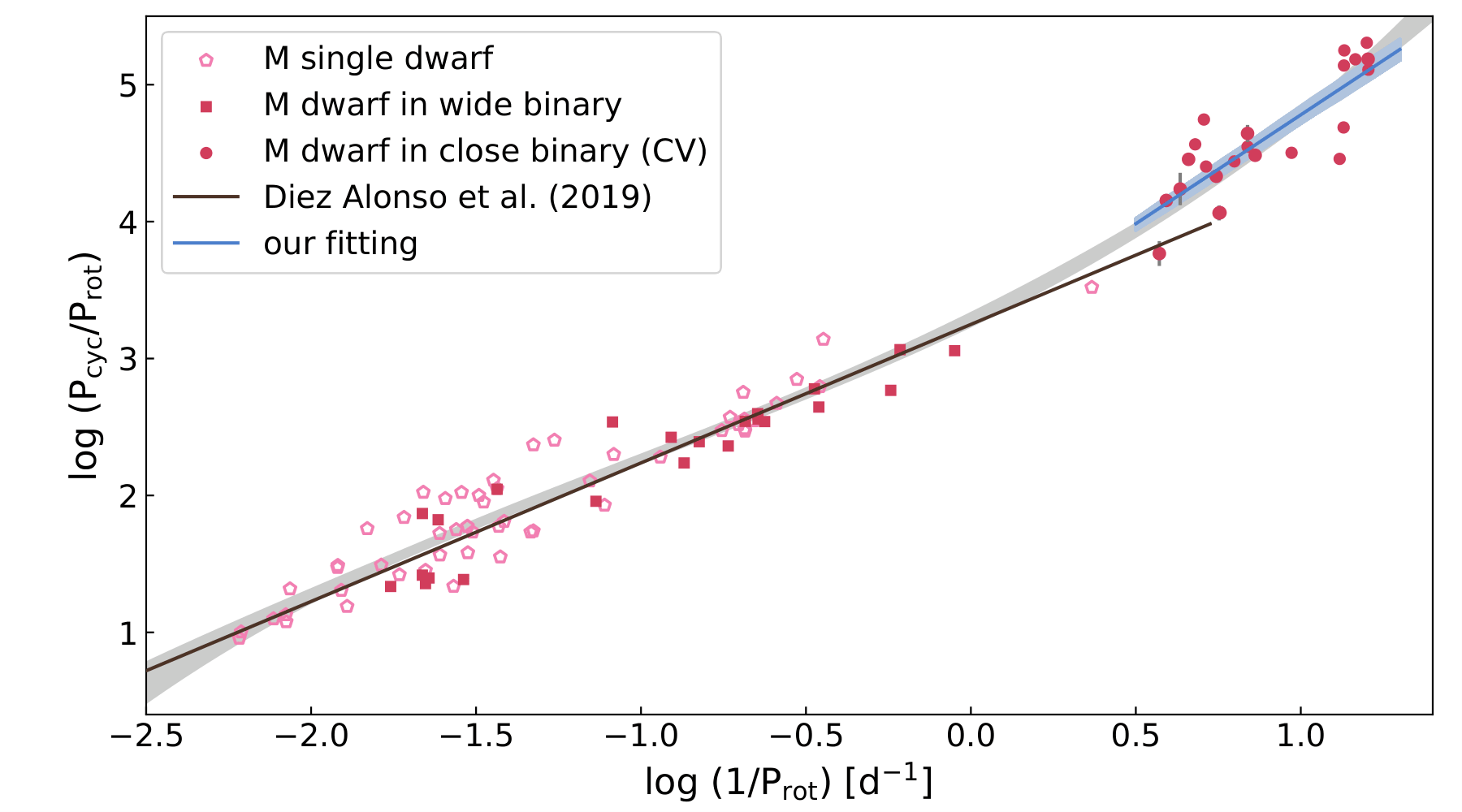}
    \caption{$P_{\rm cyc}$/$P_{\rm rot}$ versus 1/$P_{\rm rot}$ in log--log scale for M stars. Open pentagons represent single stars. Filled dots represent stars in close binaries (i.e., CVs), while filled squares represent stars in wide binaries. The solid black line is the fitting from literature \cite{2019A&A...621A.126D} for single M dwarfs with a slope of 1.01$\pm$0.06. The blue line represents our fitting using M dwarfs in close binaries, with a slope of 1.59$\pm$0.05, and the shaded blue region indicates the 1$\sigma$ uncertainty of the linear fit. The shaded gray region denotes the 1$\sigma$ uncertainty of the polynomial fit to the whole sample.
    }
    \label{figure2.fig}
\end{figure*}

Previous studies have reported no correlation between $P_{\rm cyc}$ and $P_{\rm rot}$ for single M dwarfs. 
The derived slopes for the log($P_{\rm cyc}$/$P_{\rm rot}$) versus log(1/$P_{\rm rot}$) 
relationship in these studies include: 1.03$\pm$0.04 \cite{2012ARep...56..716S}, 1.10$\pm$0.03 \cite{2018A&A...612A..89S}, and 1.01$\pm$0.06 \cite{2019A&A...621A.126D}, etc. 
In contrast, our analysis of M dwarfs in CV systems yields a significantly steeper slope.
The best-fit linear relation between log($P_{\rm cyc}$/$P_{\rm rot}$) and log($1/P_{\rm rot}$) for these M-star CVs (Figure \ref{figure2.fig}) yields
\begin{equation}
    {\rm log}(P_{\rm cyc}/P_{\rm rot}) = {(1.59\pm0.05)} \times {\rm log}(1/P_{\rm rot}) + (3.19\pm0.04).
\end{equation}
This reveals a clear connection between $P_{\rm cyc}$ and $P_{\rm rot}$: 
\begin{equation}
    P_{\rm cyc} \propto P_{\rm rot}^{-0.59},
\end{equation}
suggesting faster-rotating stars tend to maintain longer cycle periods (see Figure \ref{figure1.fig}).

We also performed a third-order polynomial fit to the relation between log($P_{\rm cyc}$/$P_{\rm rot}$) and log($1/P_{\rm rot}$) for the whole sample:
\begin{equation}
\begin{aligned}
    {\rm log}(P_{\rm cyc}/P_{\rm rot}) = (3.284 \pm 0.049) + (1.173 \pm 0.042)x  \\
   + (0.230 \pm 0.050)x^2 + (0.074 \pm 0.028)x^3,
    \end{aligned}
\end{equation}
where $x =$ log($1/P_{\rm rot}$). This fit clearly reveals variations in the relation across different stellar systems.
The contrasting behavior between M dwarfs in close binaries and single M dwarfs suggests differences in their underlying dynamo mechanisms.

\section{Simulation}
\label{sec:simulation}

In close binary systems, stars cannot be treated as spherical fluids as single stars, due to the asymmetric deformation caused by rotation and tidal forces from their companions.
This deformation not only redistributes stellar mass but also changes internal physical fields and thermal status of the star, which may influence magnetic dynamo processes and be reflected in the observed relation between $P_{\rm cyc}$ and $P_{\rm rot}$.
Given that stars in close binaries exhibit longer activity cycle periods, a plausible explanation is that their dynamo actions are altered, requiring longer timescales to accumulate sufficient energy to drive a magnetic cycle.

\begin{figure}[H]
	\includegraphics[width=0.48\textwidth]{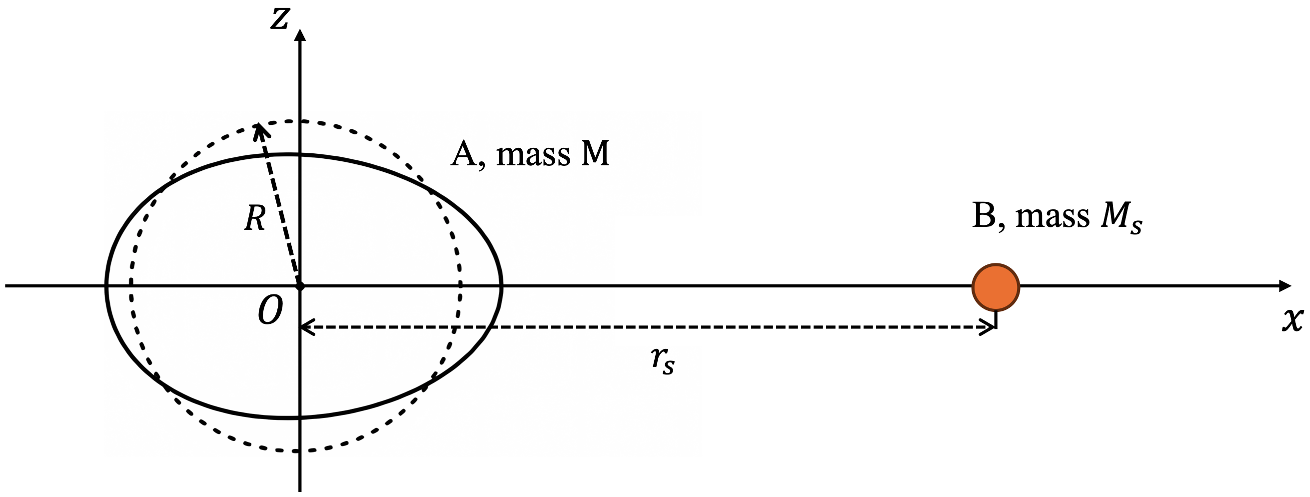}
	\caption{{Sketch of a binary system under the asymmetrical deformation.} Star A of mass $M$ and the companion B of mass $m_s$ are tidally locked to each other. The dashed circle denotes the original spherically symmetric shape of star A of radius $R$, while the solid curve represents its asymmetrically deformed figure due to its fast rotation and the tidal forcing from the companion B. In this physical model the companion B is simplified as a mass point. The distance between the barycenters of A and B is $r_s$.}
	\label{figure3.fig}
\end{figure}

We constructed a binary model to study the effects of the asymmetrical deformation of the figure of stars, due to rotation and tidal forcing from their companions, on thermal convection inside them (Figure \ref{figure3.fig}). In this physical model, star A of mass $M$ is tidally locked to its companion, star B, which is simplified as a point mass $m_s$. The origin $O$ of the coordinate system is set at the center of mass of star A, while the mass point B is located on the $x$-axis at $\boldsymbol{r}_s = r_s\boldsymbol{\hat{x}}$. Star A rotates along the $z$-axis with a constant angular velocity $\boldsymbol{\Omega}=\Omega\boldsymbol{\hat{z}}$. The $y$-axis is perpendicular to the $xOz$ plane pointing inward. To focus on the effects of asymmetric stellar figures on thermal convection, star A was modelled by an internally heated Boussinesq fluid of constant density $\rho_0$ by constant heat sources $Q_h$.  The material properties of star A, such as the kinematic viscosity $\nu$, the thermal diffusivity $\kappa$, the thermal expansion coefficient $\alpha_T$, and the specific heat capacity of constant pressure $c_p$, are assumed to be constant as well. The dynamics of thermal convection in star A is then governed by the following dimensional equations in the co-rotating frame of reference \cite{chandrasekhar1961}:
\begin{equation}
\begin{aligned}
	\rho_0 \biggl( \frac{\partial\boldsymbol{u}}{\partial t} + \boldsymbol{u}\cdot\nabla\boldsymbol{u} + 2\boldsymbol{\Omega}\times\boldsymbol{u} \biggr) = -\nabla p + \rho [\boldsymbol{g}_0 - \boldsymbol{\Omega} \times (\boldsymbol{\Omega} \times \boldsymbol{r}) \\ + \boldsymbol{g}_T]
    + \rho_0 \nu \nabla^2\boldsymbol{u},
    \end{aligned}
    \end{equation}
    
    \begin{equation}
	\frac{\partial T}{\partial t} + \boldsymbol{u} \cdot \nabla T = \kappa \nabla^2 T + \frac{Q_h}{c_p \rho_0},\ {\rm and}
    \end{equation}
    
    \begin{equation}
	\nabla\cdot\boldsymbol{u} = 0,
\end{equation}
where $\boldsymbol{u}$ is the velocity of a fluid parcel marked by a position vector $\boldsymbol{r}$, $p$ represents the hydrodynamic pressure, and $T$ denotes the temperature. The density related to the buoyancy is determined as $\rho = \rho_0 (1-\alpha_T T)$. Besides the self-gravity $\boldsymbol{g}_0$, the contributions of the centrifugal force $-\boldsymbol{\Omega}\times (\boldsymbol{\Omega}\times \boldsymbol{r})$ due to rotation and the tidal force $\boldsymbol{g}_T$ imposed by the mass point B  are also considered in the buoyancy in this physical model. The no-slip boundary condition for velocity, the isothermal boundary condition for temperature and the equipotential equilibrium condition 
\begin{equation}
	\boldsymbol{u} = \boldsymbol{0},\quad T = 0,\quad U = \text{constant},
\end{equation}
were adopted at the boundary surface of star A. Here, the total potential $U$ is the sum of the self-gravity potential, the centrifugal potential, and the tide-generating potential.

The key step to study thermal convection within star A is the determination of its deformed figure shaped by the rotation and tidal forcing imposed by the mass point B. As star A is tidally locked to its companion B, the figure of A is taken to be that
in the hydrostatic state in the co-rotating frame of reference. The non-perturbative iteration method is employed to obtain the time-independent figure of star A in the hydrostatic state \cite{Hubbard2012,Hubbard2013,Wahl2017,Wahl2020}. The accuracy of this iteration method has been investigated \cite{Hubbard2014}. According to the iteration method, the deformed figure is controlled by the two dimensionless parameters $q_{\text{rot}} = \frac{\Omega^2R^3}{GM}$ and $q_{\text{tid}}=-\frac{3m_sR^3}{Mr_s^3}$, where $G$ is the universal gravitational constant and $R$ denotes the supposed spherical radius of  star A when  getting rid of rotation and tides (i.e., the radius of the dashed circle in Figure \ref{figure3.fig}). 
As the Keplerian orbit is adopted in this tide-locked system, only $q_{\text{rot}}$ is free, which can be re-expressed as
\begin{equation}
	q_{\text{rot}} = \biggl(1+\frac{m_s}{M}\biggr) \biggl( \frac{R}{r_s} \biggr)^3.
\end{equation} 
As the mass $m_s$ of the companion B becomes relatively larger than that of star A or the ratio of the spherical radius $R$ to the distance $r_s$ between these two objects becomes larger, the value of $q_{\text{rot}}$ increases, resulting in a more heavily deformed figure of star A. The volume covered by the deformed surface calculated by the iteration method for any given $q_{\text{rot}}$ is required to be the same as that of the sphere of radius $R$ such that the conservation of mass of star A of constant density is guaranteed. 

Once the deformed figures of star A are obtained, direct numerical simulations of thermal convection were performed using the finite element method. The numerical method and the overall code have been successfully used and validated in modeling of thermal convection in rotating oblate spheroids flattened by the effects of fast rotation alone \cite{kong2022,li2022,li2023,li2024}. It's clear that the dynamics of thermal convection in star A is characterized by the Ekman number $\Ek$ which measures the relative importance of the viscous force and the Coriolis force, the Prandtl number $\Prt$ which is the ratio of the kinematic viscosity $\nu$ to the thermal diffusivity $\kappa$, the Rayleigh number $\Ra$ which measures the ratio of destabilizing buoyancy to stabilizing processes, and the value of $q_{\text{rot}}$ which decides the deformed figure of star A at hydrostatic state in the rotating frame of reference. The thermal convection of stars is in the regime of inertial convection, which is marked by the asymptotically small Ekman number $\Ek \ll 1$ and the sufficiently small Prandtl number $\Prt < 1$. In all the numerical simulation cases, the Ekman number and the Prandtl number are fixed at $\Ek=10^{-4}$ and $\Prt=5\times 10^{-3}$ (for computational feasibility), respectively, in order to investigate the effects of asymmetrical deformation of different degree, measured quantitatively by values of $q_{\text{rot}}$, on thermal convection in star A. Since the velocity decreases to zero in the very thin boundary layer, the computational domain is discretized using an unstructured tetrahedral mesh of second order with a typical fine mesh size $\mathcal{O}$($R\sqrt{\Ek}$) near the boundary surface. More specifically, in the thin region of depth $0.02R$ adjacent to the boundary, the max size of mesh is set to be below $R\sqrt{\Ek}$.

\begin{figure*}
	\includegraphics[width=0.95\textwidth]{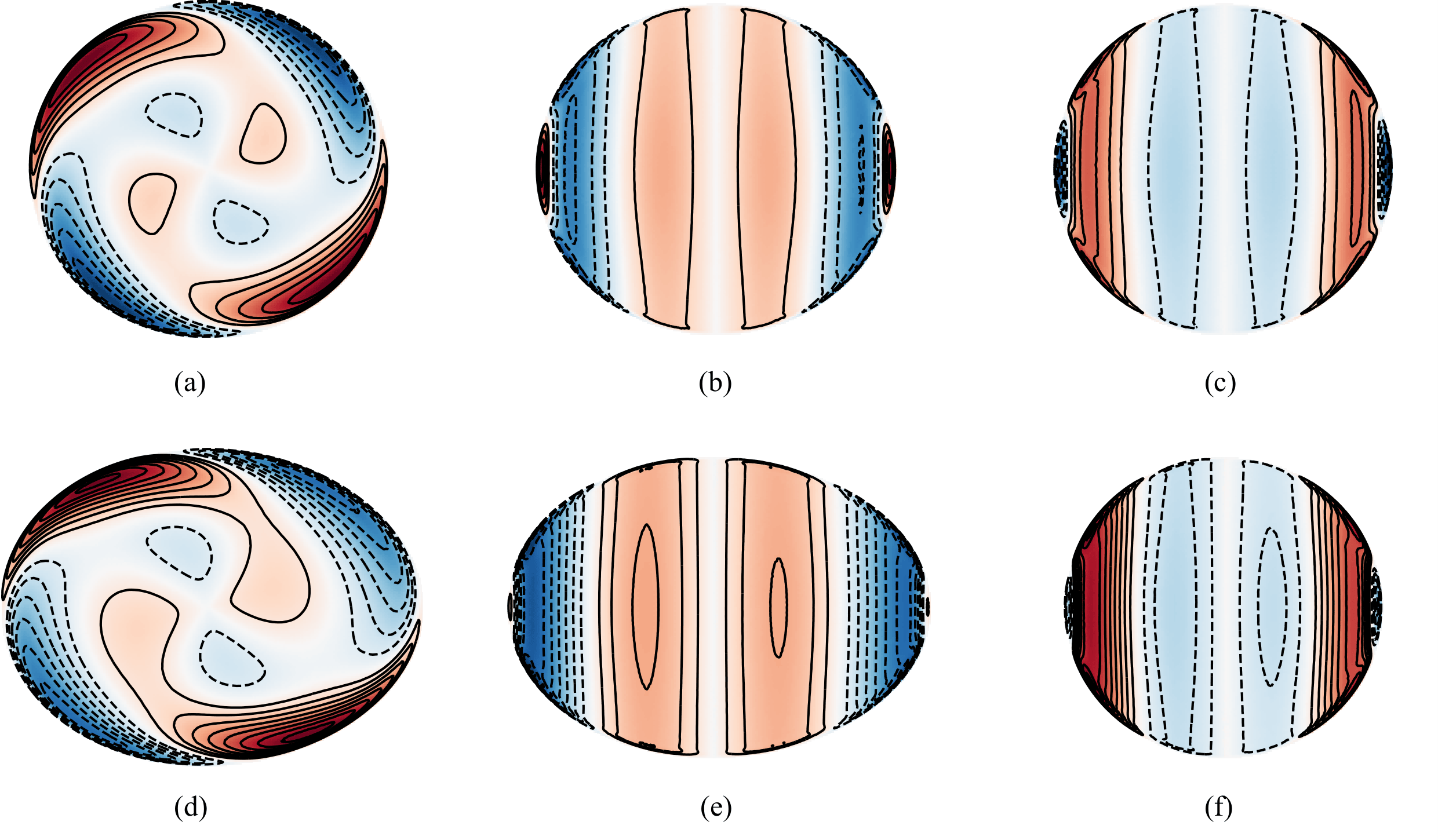}
	\caption{Contours of the azimuthal component $\boldsymbol{\hat{\phi}}\cdot\boldsymbol{u}$ of flows at the onset of thermal convection computed via direct numerical simulation. The top row represents the numerical results in the slightly deformed figure with $q_{\text{rot}}=0.015$, while the bottom row demonstrates the numerical results in the heavily deformed figure with $q_{\text{rot}}=0.051$. The panels (a) and (d) are sliced in the $xOy$ plane, (b) and (e) in the $xOz$ plane, and (c) and (f) in the $yOz$ plane. Other physical parameters are referred to Table \ref{table1.tab}.}
    \label{figure4.fig}
\end{figure*}

We conducted numerical simulations for two sets of physical configurations: one representing  a slightly deformed spherical-like figure with $m_s/M=50$ and $r_s/R=15$, and the other  representing a significantly deformed asymmetrical figure with $m_s/M=50$ and $r_s/R=10$. Taking the critical Rayleigh number $Ra_c=1.63$ of thermal convection within the undeformed sphere as the baseline \cite{li2022}, we found that $\Ra_c$  increases by $6\%$ to $1.74$ for the spherical-like figure with $q_{\text{rot}}=0.015$, while by $31\%$ to $2.13$ for the heavily deformed figure with $q_{\text{rot}}=0.051$ (Table \ref{table1.tab}). This increasing tendency of $Ra_c$ with $q_{\text{rot}}$ shows a stronger repression of thermal convection as a star undergoes heavier asymmetrical deformation caused by rotation and tidal forces. It is also indicated (in Figure \ref{figure4.fig}) that the velocity field at the onset of thermal convection becomes non-axisymmetrically stretched when the figure of a star becomes significantly deformed. It should be noted that the definition of the Rayleigh number $Ra$ here is identical to that of the stratification number $\St$ in \cite{li2022}.

\begin{figure*}
    \includegraphics[width=0.95\textwidth]{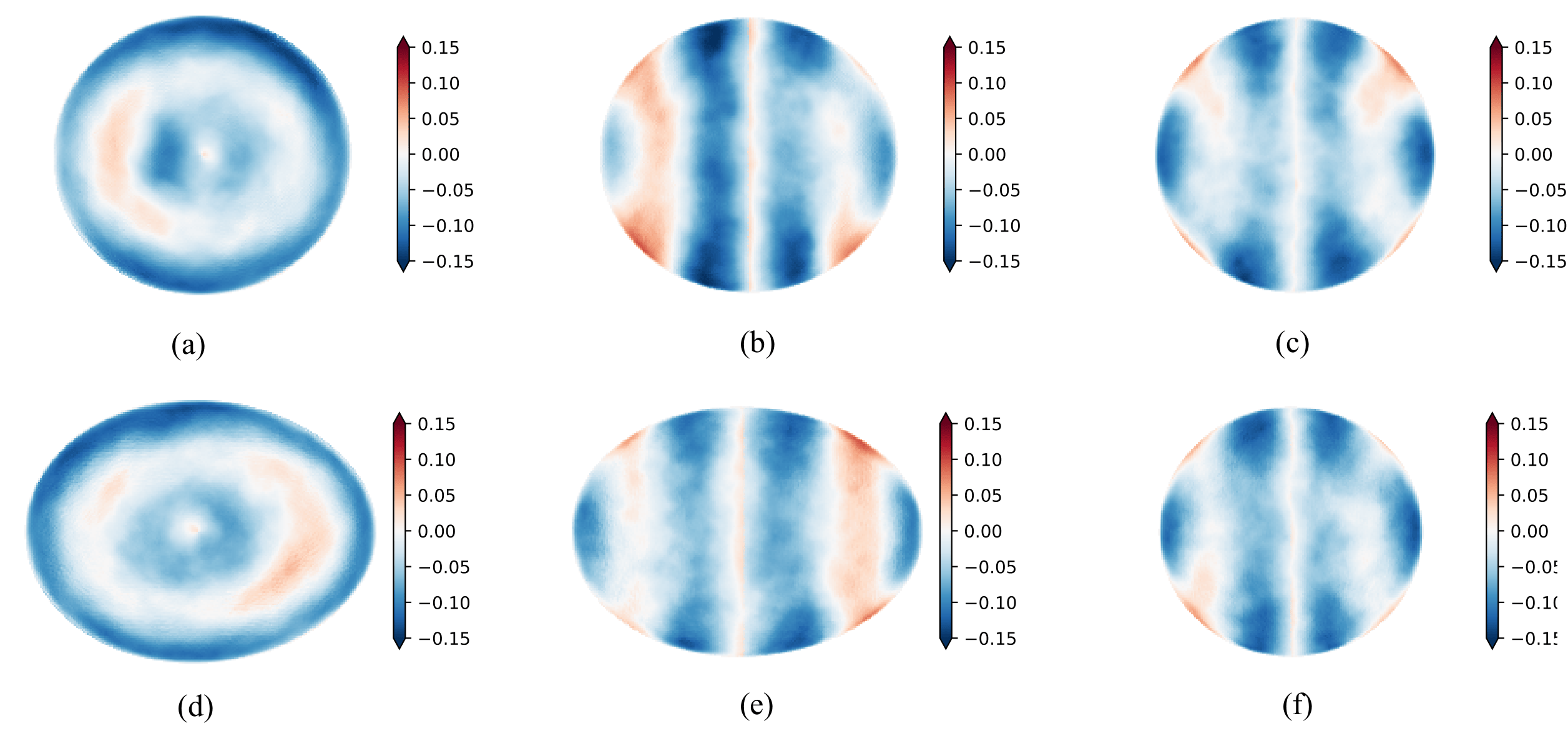}
	\caption{Contours of the time-averaged azimuthal velocity $\boldsymbol{\hat{\phi}}\cdot\boldsymbol{\widebar{U}}$ of turbulent flows via simulation. The top panels represent the numerical results in the slightly deformed figure with $q_{\text{rot}}=0.015$, while the bottom panels describe the numerical results in the heavily deformed figure with $q_{\text{rot}}=0.051$. The unit of velocity is $R\Omega$, with $R$ and $\Omega$ being the spherical radius and strength of angular velocity of rotation, respectively. The Rayleigh number is fixed at $\Ra=10\Ra_c$, respetively, in these numerical results. Other physical parameters are referred to Table \ref{table1.tab}. The panels (a) and (d) are sliced in the $xOy$ plane, (b) and (e) in the $xOz$ plane, and (c) and (f) in the $yOz$ plane.}
    \label{figure5.fig}
\end{figure*}

\begin{figure}[H]
	\includegraphics[width=0.48\textwidth]{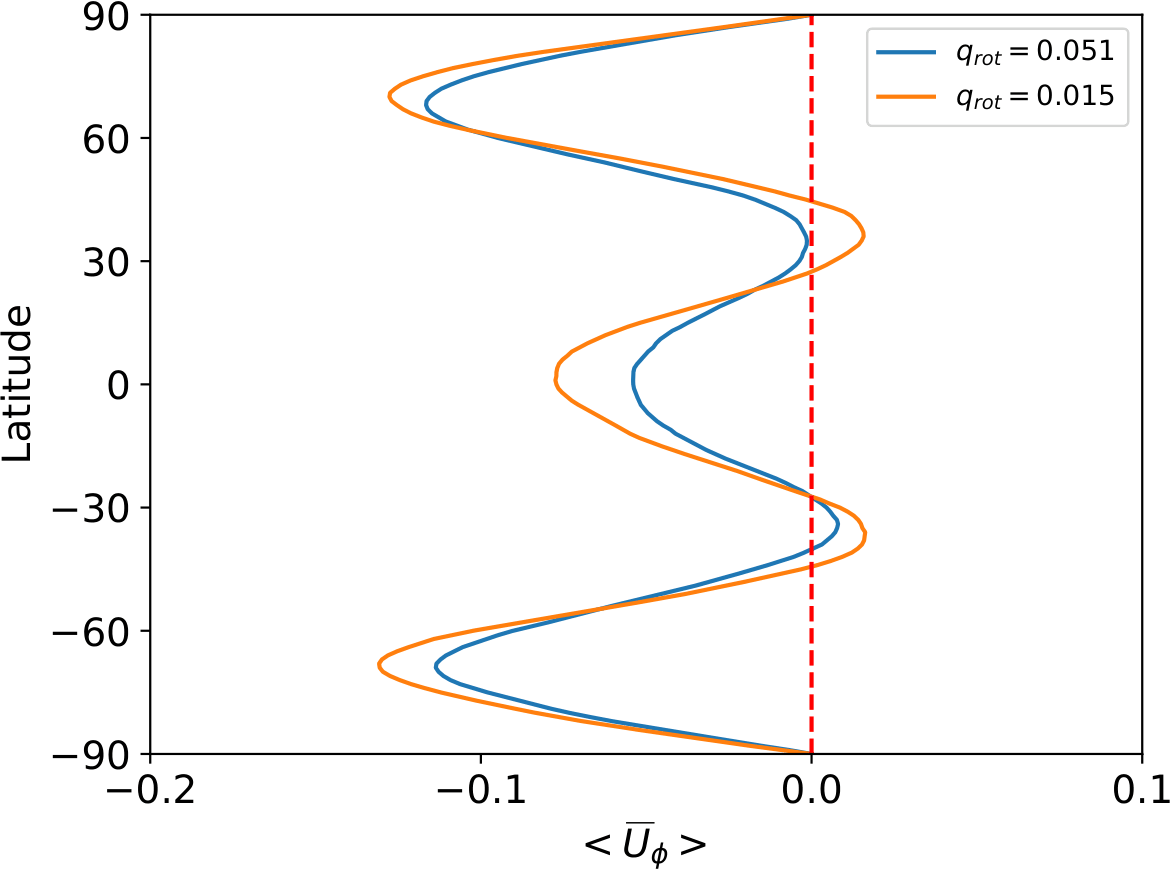}
	\caption{{Mean zonal flows at the spherical surface of radius $r=0.85R$ (centered at the origin) for two representative values of $q_{\text{rot}}$.} The Rayleigh number, Ekman number and Prandtl number are fixed at $\Ra=10\Ra_c$, $\Ek=10^{-4}$ and $\Prt=5\times 10^{-3}$, respectively.
    It can be seen that the mean zonal flows tend to weaken as stellar deformation increases.
    }
    \label{figure6.fig}
\end{figure}

Since the critical Rayleigh number $\Ra_c$ increases as the figure undergoes heavier asymmetric deformation, it is revealed (in Figure \ref{figure5.fig}) that the strength of time-averaged thermal convection in stars of close binaries with significantly deformed figures becomes slightly weaker than that in spherical-like stars at the same level of supercritical convection (e.g., at $\Ra=10\Ra_c$). 
The total kinetic energy, dimensionalized by the unit $\rho_0R^5\Omega^2$, of the time-averaged azimuthal velocity $\boldsymbol{\hat{\phi}}\cdot\boldsymbol{\widebar{U}}$ normalized by $R\Omega$ decreases slightly from $0.0068$ at $q_{\text{rot}}=0.015$ to $0.0060$ at $q_{\text{rot}}=0.051$. Here $\rho_0$ is the density in the hydrostatic state, $\boldsymbol{\widebar{U}}$ represents the time-averaged velocity, and $\boldsymbol{\hat{\phi}}$ denotes the conventional longitudinal direction in the polar spherical coordinates.
Subsequently, as shown in Figure \ref{figure6.fig}, the mean zonal flows $\left<\widebar{U}\right>$ averaged from the results of Figure \ref{figure5.fig} over the longitude at the spherical surface of radius $r=0.85R$, simulated at $\Ra=10\Ra_c$ with two different values of $q_{\text{rot}}$, also indicate that the differential rotation tends to weaken as stellar deformation increases.

\section{Discussion and conclusions}
\label{sec:discussion}

We compiled a sample of M dwarfs as single stars or components of binary systems.
These stars, spanning a range of tidal distortions, were used to investigate the relationship between stellar cycle periods ($P_{\rm cyc}$) and rotation periods ($P_{\rm rot}$), which are key observables linked to stellar dynamo processes.

From our analysis of tidally distorted M dwarfs in CVs, we identified a strong anti-correlation between $P_{\rm cyc}$ and $P_{\rm rot}$, characterized by ${\rm log}(P_{\rm cyc}/P_{\rm rot}) \propto ({\rm log}(1/P_{\rm rot}))^{1.59\pm0.05}$. This contrasts with the nearly linear relation found in single M dwarfs (slope $\approx$ 1.01--1.10). Our result means that stars experiencing stronger tidal distortion tend to have longer magnetic cycles, suggesting modifications to the dynamo mechanism under non-spherical geometries.

To understand the physical origin of the varying $P_{\rm cyc}$–$P_{\rm rot}$ relation, we constructed a binary model to study the influence of asymmetric stellar deformation on thermal convection. Our simulations show that thermal convection weakens in highly distorted stars, and subsequently, the differential rotation is also reduced. These, in turn, may increase the timescale for magnetic field conversion between poloidal and toroidal components, offering a plausible explanation for the newly identified $P_{\rm cyc}$–$P_{\rm rot}$ relation observed in CVs.

\begin{figure}[H]
	\includegraphics[width=0.48\textwidth]{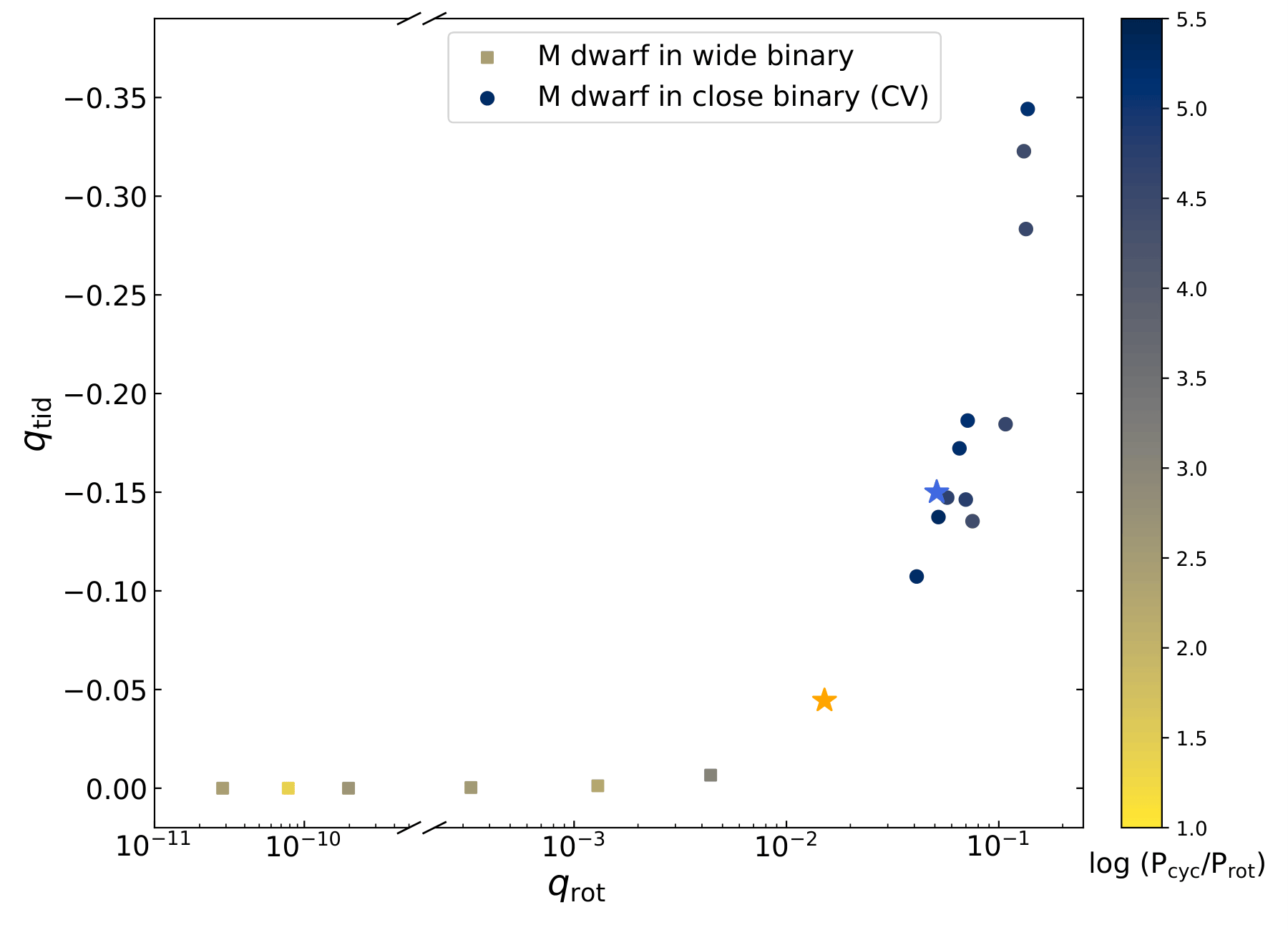}
	\caption{{Distribution of the M binary sample in the $q_{\rm tid}$ versus $q_{\rm rot}$ diagram.} The orange and blue asterisks represent the two simulations with $q_{\rm rot} =$ 0.015 and 0.051, respectively. The colorbar represents the log($P_{\rm cyc}$/$P_{\rm rot}$) values.
    }
    \label{figure7.fig}
\end{figure}

We calculated the dimensionless parameters $q_{\rm rot}$ and $q_{\rm tid}$ for our M-star sample, by compiling stellar and orbital parameters for wide and close binaries from previous studies, including stellar types, component masses, orbital periods, etc (see Tables S4 and S5 for details).
As in our simulation, $q_{\rm rot}$ and $q_{\rm tid}$ were calculated following $q_{\text{rot}} = (1+\frac{m_s}{M})(\frac{R}{r_s})^3$ and $q_{\text{tid}}=-\frac{3m_sR^3}{Mr_s^3}$, respectively. Here $M$ represents the mass of the observed star, corresponding to the brighter component in normal binaries or the non-compact star (i.e., the M dwarf) in binaries with compact companions. 
$R$ is the spherical radius of the observed star when getting rid of rotation and tides, estimated with $R/R_{\odot} \approx (M/M_{\odot})^{0.9}$ \cite{1984ApJS...54..443P}.
$m_s$ is the mass of the companion star.
The binary separation $r_s$ was derived by Kepler's third law: ${r_s}^3=(M+m_s)P_{\rm orb}^2$. 
Figure \ref{figure7.fig} shows the distribution of our binary sample in the $q_{\rm tid}$ versus $q_{\rm rot}$ diagram, compared to the two sets of simulations.
Most M dwarfs in our CV sample exhibit deformation levels exceeding that represented by the simulation with $q_{\rm rot} =$ 0.051.
In addition, as an example, Figure S1 presents mesh plots for the wide binary GJ 1054 A and the close binary DQ Her, which clearly illustrates distinct stellar geometries.

In the Parker kinematic migratory dynamo model \cite{1955ApJ...122..293P}, the cycle period $P_{\rm cyc}$ is proportional to the dynamo number $D$ by $P_{\rm cyc} \sim D^{-2/3}$  \cite{1996ApJ...460..848B}.
The dynamo number is given by $D = \alpha \Delta\Omega d^3/\eta^2 \sim \alpha \Delta\Omega$  \cite{1993ApJ...414L..33S}, where
$d$ is the characteristic convection length scale,
$\eta$ is the turbulent magnetic diffusivity,
$\alpha$ represents the helicity, 
$\Omega$ is the angular velocity, 
and $\Delta\Omega$ denotes differential rotation. 
Thus, the cycle period $P_{\mathrm{cyc}}$ scales as $\left[ \eta^2/(\alpha \Delta\Omega) \right]^{3/2}$.

Our simulations indicated that deformed stars exhibit suppressed thermal convection. This suppression has two competing effects on the dynamo process. On one hand, it reduces the kinematic helicity $\alpha$ (which originates from the convection--rotation interaction) and weakens the mean zonal flows, implying a decrease in the differential rotation shear $\Delta\Omega$. Both observations \cite{2005MNRAS.357L...1B} and theoretical \cite{2011MNRAS.411.1059K} studies have reported that low-mass stars with weaker convection possess reduced differential rotation. On the other hand, the turbulent magnetic diffusivity $\eta$ is also expected to decrease in such stars, which would enhance the dynamo efficiency in generating magnetic field.
Consequently, the resulting cycle period depends on which of these two effects dominates: the reduction of $\alpha$ and $\Delta\Omega$ versus the decrease in $\eta$. In this study, the observed new $P_{\rm cyc}$--$P_{\rm rot}$ relation, together with our simulations, suggests that the reduction in $\alpha$ and $\Delta\Omega$ is the dominant process, leading to a lengthening of the activity cycle.

Additional observational evidences also suggest the link between dynamo processes and stellar deformation.
For example, magnetically active close binaries (which, we note, are not classified as CVs in this work), often exhibit starspots concentrated at preferred longitudes  \cite{1994A&A...291..110O,1995ApJS...97..513H,1998AJ....115.1145H,2002A&A...389..202O}, which can be explained by the formation of clusters of flux tube eruptions at specific longitudes, driven by tidal forces and deviations of stellar structure from spherical symmetry  \cite{2003A&A...405..291H,2003A&A...405..303H}.
Moreover, such a connection between dynamo retardment and stellar deformation may help explain the well-known supersaturation phenomenon, where stellar activity decreases with increasing rotational velocity, as observed in X-ray, chromospheric, and photometric emissions  \cite{2011ApJ...743...48W, 2022MNRAS.514.4932C, 2024ApJ...976..243D, 2025NatAs...9.1045Y}, particularly in binary systems. 
Close binary systems thus provide a cosmic laboratory to probe dynamo mechanisms under extreme stellar geometries.

\begin{table}[H]
	\caption{Critical Rayleigh numbers in different physical configurations. \label{table1.tab}}
	\begin{center}
	\begin{tabular}{ccccll}
		\hline
		$\Ek$ & $\Prt$ & $m_s/M$ & $r_s/R$ & $q_{\text{rot}}$ & $\Ra_c$\\ \hline
		            &               & ---             & ---           & $0$                       & $1.63^{\ast}$ \\
		 $10^{-4}$ & $5\times 10^{-3}$ & $50$ & $15$ & $0.015$ & $1.74$ \\
		                       &                                     & $50$  & $10$ & $0.051$ & $2.13$\\ \hline
	\end{tabular}
\end{center}
	$^{\ast} $The critical Rayleigh number $\Ra_c=1.63$ for rotating spheres at the given $\Ek$ and $\Prt$ is  referred to \cite{li2022}, in which the Rayleigh number is denoted as the stratification number. 
\end{table}

\Acknowledgements{J.F.L. was funded by the National Natural Science Foundation of China (NSFC) under grant number 12588202. 
S.W. was funded by the National Key Research and Development Program of China under grant number 2023YFA1607901, the Strategic Priority Program of the Chinese Academy of Sciences under grant number XDB1160302, the NSFC under grant number 12273057, and science research grants from the China Manned Space Project. 
W.B.L. was funded by the NSFC under grant number 12403070 and the Shanghai Post-doctoral Excellence Program under number 2023000137.
D.L.K. was funded by the NSFC under grant numbers 12425306 and 12250013.
J.F.L. was also supported by the New Cornerstone Science Foundation through the New Cornerstone Investigator Program and the XPLORER PRIZE.}

\InterestConflict{The authors declare that they have no conflict of interest.}




\bibliography{scibib}

@ARTICLE{1995ApJS...97..513H,
       author = {{Henry}, Gregory W. and {Eaton}, Joel A. and {Hamer}, Jamesia and {Hall}, Douglas S.},
        title = "{Starspot Evolution, Differential Rotation, and Magnetic Cycles in the Chromospherically Active Binaries lambda Andromedae, sigma Geminorum, II Pegasi, and V711 Tauri}",
      journal = {\apjs},
         year = 1995,
        month = apr,
       volume = {97},
        pages = {513},
          doi = {10.1086/192149},
       adsurl = {https://ui.adsabs.harvard.edu/abs/1995ApJS...97..513H}
}

@ARTICLE{2002A&A...389..202O,
       author = {{Ol{\'a}h}, K. and {Strassmeier}, K.~G. and {Weber}, M.},
        title = "{Doppler imaging of stellar surface structure. XVIII. The very active RS CVn binary UZ Librae revisited}",
      journal = {\aap},
         year = 2002,
        month = jul,
       volume = {389},
        pages = {202-212},
          doi = {10.1051/0004-6361:20020506},
       adsurl = {https://ui.adsabs.harvard.edu/abs/2002A&A...389..202O}
}

@ARTICLE{1994A&A...291..110O,
       author = {{Olah}, K. and {Budding}, E. and {Kim}, H. -L. and {Etzel}, P.~B.},
        title = "{The active close binary system ER Vulpeculae.}",
      journal = {\aap},
         year = 1994,
        month = nov,
       volume = {291},
        pages = {110-120},
       adsurl = {https://ui.adsabs.harvard.edu/abs/1994A&A...291..110O}
}

@ARTICLE{1998AJ....115.1145H,
       author = {{Heckert}, Paul A. and {Maloney}, George V. and {Stewart}, Maria C. and {Ordway}, James I. and {Hickman}, Ann and {Zeilik}, Michael},
        title = "{A Decade of Starspot Activity on the Eclipsing Short-period RS Canum Venaticorum Star WY Cancri: 1988-1997}",
      journal = {\aj},
         year = 1998,
        month = mar,
       volume = {115},
       number = {3},
        pages = {1145-1152},
          doi = {10.1086/300238},
       adsurl = {https://ui.adsabs.harvard.edu/abs/1998AJ....115.1145H}
}

@ARTICLE{2022MNRAS.510.4736K,
       author = {{Knight}, Amy H. and {Ingram}, Adam and {Middleton}, Matthew and {Drake}, Jeremy},
        title = "{Eclipse mapping of EXO 0748-676: evidence for a massive neutron star}",
      journal = {\mnras},
         year = 2022,
        month = mar,
       volume = {510},
       number = {4},
        pages = {4736-4756},
          doi = {10.1093/mnras/stab3722},
archivePrefix = {arXiv},
       eprint = {2201.02188},
 primaryClass = {astro-ph.HE},
       adsurl = {https://ui.adsabs.harvard.edu/abs/2022MNRAS.510.4736K}
}

@ARTICLE{2019A&A...622A..40K,
       author = {{K{\"u}ker}, M. and {R{\"u}diger}, G. and {Olah}, K. and {Strassmeier}, K.~G.},
        title = "{Cycle period, differential rotation, and meridional flow for early M dwarf stars}",
      journal = {\aap},
         year = 2019,
        month = feb,
       volume = {622},
          eid = {A40},
        pages = {A40},
          doi = {10.1051/0004-6361/201833173},
archivePrefix = {arXiv},
       eprint = {1804.02925},
 primaryClass = {astro-ph.SR},
       adsurl = {https://ui.adsabs.harvard.edu/abs/2019A&A...622A..40K}
}

@ARTICLE{2019A&A...621A.126D,
       author = {{D{\'\i}ez Alonso}, E. and {Caballero}, J.~A. and {Montes}, D. and {de Cos Juez}, F.~J. and {Dreizler}, S. and {Dubois}, F. and {Jeffers}, S.~V. and {Lalitha}, S. and {Naves}, R. and {Reiners}, A. and {Ribas}, I. and {Vanaverbeke}, S. and {Amado}, P.~J. and {B{\'e}jar}, V.~J.~S. and {Cort{\'e}s-Contreras}, M. and {Herrero}, E. and {Hidalgo}, D. and {K{\"u}rster}, M. and {Logie}, L. and {Quirrenbach}, A. and {Rau}, S. and {Seifert}, W. and {Sch{\"o}fer}, P. and {Tal-Or}, L.},
        title = "{CARMENES input catalogue of M dwarfs. IV. New rotation periods from photometric time series}",
      journal = {\aap},
         year = 2019,
        month = jan,
       volume = {621},
          eid = {A126},
        pages = {A126},
          doi = {10.1051/0004-6361/201833316},
archivePrefix = {arXiv},
       eprint = {1810.03338},
 primaryClass = {astro-ph.SR},
       adsurl = {https://ui.adsabs.harvard.edu/abs/2019A&A...621A.126D}
}

@ARTICLE{2018A&A...616A.160V,
       author = {{Viviani}, M. and {Warnecke}, J. and {K{\"a}pyl{\"a}}, M.~J. and {K{\"a}pyl{\"a}}, P.~J. and {Olspert}, N. and {Cole-Kodikara}, E.~M. and {Lehtinen}, J.~J. and {Brandenburg}, A.},
        title = "{Transition from axi- to nonaxisymmetric dynamo modes in spherical convection models of solar-like stars}",
      journal = {\aap},
         year = 2018,
        month = aug,
       volume = {616},
          eid = {A160},
        pages = {A160},
          doi = {10.1051/0004-6361/201732191},
archivePrefix = {arXiv},
       eprint = {1710.10222},
 primaryClass = {astro-ph.SR},
       adsurl = {https://ui.adsabs.harvard.edu/abs/2018A&A...616A.160V}
}

@ARTICLE{2018A&A...612A..89S,
       author = {{Su{\'a}rez Mascare{\~n}o}, A. and {Rebolo}, R. and {Gonz{\'a}lez Hern{\'a}ndez}, J.~I. and {Toledo-Padr{\'o}n}, B. and {Perger}, M. and {Ribas}, I. and {Affer}, L. and {Micela}, G. and {Damasso}, M. and {Maldonado}, J. and {Gonz{\'a}lez-Alvarez}, E. and {Leto}, G. and {Pagano}, I. and {Scandariato}, G. and {Sozzetti}, A. and {Lanza}, A.~F. and {Malavolta}, L. and {Claudi}, R. and {Cosentino}, R. and {Desidera}, S. and {Giacobbe}, P. and {Maggio}, A. and {Rainer}, M. and {Esposito}, M. and {Benatti}, S. and {Pedani}, M. and {Morales}, J.~C. and {Herrero}, E. and {Lafarga}, M. and {Rosich}, A. and {Pinamonti}, M.},
        title = "{HADES RV programme with HARPS-N at TNG. VII. Rotation and activity of M-dwarfs from time-series high-resolution spectroscopy of chromospheric indicators}",
      journal = {\aap},
         year = 2018,
        month = apr,
       volume = {612},
          eid = {A89},
        pages = {A89},
          doi = {10.1051/0004-6361/201732143},
archivePrefix = {arXiv},
       eprint = {1712.07375},
 primaryClass = {astro-ph.SR},
       adsurl = {https://ui.adsabs.harvard.edu/abs/2018A&A...612A..89S}
}

@ARTICLE{2017A&A...606A..58D,
       author = {{Distefano}, E. and {Lanzafame}, A.~C. and {Lanza}, A.~F. and {Messina}, S. and {Spada}, F.},
        title = "{Activity cycles in members of young loose stellar associations}",
      journal = {\aap},
         year = 2017,
        month = oct,
       volume = {606},
          eid = {A58},
        pages = {A58},
          doi = {10.1051/0004-6361/201730967},
archivePrefix = {arXiv},
       eprint = {1706.07938},
 primaryClass = {astro-ph.SR},
       adsurl = {https://ui.adsabs.harvard.edu/abs/2017A&A...606A..58D}
}

@ARTICLE{2017ApJ...845...79B,
       author = {{Brandenburg}, Axel and {Mathur}, Savita and {Metcalfe}, Travis S.},
        title = "{Evolution of Co-existing Long and Short Period Stellar Activity Cycles}",
      journal = {\apj},
         year = 2017,
        month = aug,
       volume = {845},
       number = {1},
          eid = {79},
        pages = {79},
          doi = {10.3847/1538-4357/aa7cfa},
archivePrefix = {arXiv},
       eprint = {1704.09009},
 primaryClass = {astro-ph.SR},
       adsurl = {https://ui.adsabs.harvard.edu/abs/2017ApJ...845...79B}
}

@ARTICLE{2017Sci...357..185S,
       author = {{Strugarek}, A. and {Beaudoin}, P. and {Charbonneau}, P. and {Brun}, A.~S. and {do Nascimento}, J. -D.},
        title = "{Reconciling solar and stellar magnetic cycles with nonlinear dynamo simulations}",
      journal = {Science},
         year = 2017,
        month = jul,
       volume = {357},
       number = {6347},
        pages = {185-187},
          doi = {10.1126/science.aal3999},
archivePrefix = {arXiv},
       eprint = {1707.04335},
 primaryClass = {astro-ph.SR},
       adsurl = {https://ui.adsabs.harvard.edu/abs/2017Sci...357..185S}
}

@ARTICLE{2017MNRAS.464.3281W,
       author = {{Wargelin}, B.~J. and {Saar}, S.~H. and {Pojma{\'n}ski}, G. and {Drake}, J.~J. and {Kashyap}, V.~L.},
        title = "{Optical, UV, and X-ray evidence for a 7-yr stellar cycle in Proxima Centauri}",
      journal = {\mnras},
         year = 2017,
        month = jan,
       volume = {464},
       number = {3},
        pages = {3281-3296},
          doi = {10.1093/mnras/stw2570},
archivePrefix = {arXiv},
       eprint = {1610.03447},
 primaryClass = {astro-ph.SR},
       adsurl = {https://ui.adsabs.harvard.edu/abs/2017MNRAS.464.3281W}
}

@ARTICLE{2016MNRAS.463.1342S,
       author = {{{\v{S}}imon}, Vojt{\v{e}}ch},
        title = "{Properties of the long-term optical activity of the prototype polar AM Herculis}",
      journal = {\mnras},
         year = 2016,
        month = dec,
       volume = {463},
       number = {2},
        pages = {1342-1351},
          doi = {10.1093/mnras/stw1964},
       adsurl = {https://ui.adsabs.harvard.edu/abs/2016MNRAS.463.1342S}
}

@ARTICLE{2016A&A...595A..12S,
       author = {{Su{\'a}rez Mascare{\~n}o}, A. and {Rebolo}, R. and {Gonz{\'a}lez Hern{\'a}ndez}, J.~I.},
        title = "{Magnetic cycles and rotation periods of late-type stars from photometric time series}",
      journal = {\aap},
         year = 2016,
        month = oct,
       volume = {595},
          eid = {A12},
        pages = {A12},
          doi = {10.1051/0004-6361/201628586},
archivePrefix = {arXiv},
       eprint = {1607.03049},
 primaryClass = {astro-ph.SR},
       adsurl = {https://ui.adsabs.harvard.edu/abs/2016A&A...595A..12S}
}

@ARTICLE{2016A&A...591A..43D,
       author = {{Distefano}, E. and {Lanzafame}, A.~C. and {Lanza}, A.~F. and {Messina}, S. and {Spada}, F.},
        title = "{Lower limit for differential rotation in members of young loose stellar associations}",
      journal = {\aap},
         year = 2016,
        month = jun,
       volume = {591},
          eid = {A43},
        pages = {A43},
          doi = {10.1051/0004-6361/201527698},
archivePrefix = {arXiv},
       eprint = {1604.01917},
 primaryClass = {astro-ph.SR},
       adsurl = {https://ui.adsabs.harvard.edu/abs/2016A&A...591A..43D}
}

@ARTICLE{2015A&A...583A.134F,
       author = {{Ferreira Lopes}, C.~E. and {Le{\~a}o}, I.~C. and {de Freitas}, D.~B. and {Canto Martins}, B.~L. and {Catelan}, M. and {De Medeiros}, J.~R.},
        title = "{Stellar cycles from photometric data: CoRoT stars}",
      journal = {\aap},
         year = 2015,
        month = nov,
       volume = {583},
          eid = {A134},
        pages = {A134},
          doi = {10.1051/0004-6361/201424900},
archivePrefix = {arXiv},
       eprint = {1508.06194},
 primaryClass = {astro-ph.SR},
       adsurl = {https://ui.adsabs.harvard.edu/abs/2015A&A...583A.134F}
}

@ARTICLE{2014MNRAS.441.2744V,
       author = {{Vida}, K. and {Ol{\'a}h}, K. and {Szab{\'o}}, R.},
        title = "{Looking for activity cycles in late-type Kepler stars using time-frequency analysis}",
      journal = {\mnras},
         year = 2014,
        month = jul,
       volume = {441},
       number = {3},
        pages = {2744-2753},
          doi = {10.1093/mnras/stu760},
archivePrefix = {arXiv},
       eprint = {1404.4359},
 primaryClass = {astro-ph.SR},
       adsurl = {https://ui.adsabs.harvard.edu/abs/2014MNRAS.441.2744V}
}

@ARTICLE{2013AN....334..972V,
       author = {{Vida}, K. and {Kriskovics}, L. and {Ol{\'a}h}, K.},
        title = "{A quest for activity cycles in low-mass stars}",
      journal = {Astronomische Nachrichten},
         year = 2013,
        month = nov,
       volume = {334},
       number = {9},
        pages = {972},
          doi = {10.1002/asna.201211973},
archivePrefix = {arXiv},
       eprint = {1306.6845},
 primaryClass = {astro-ph.SR},
       adsurl = {https://ui.adsabs.harvard.edu/abs/2013AN....334..972V}
}

@ARTICLE{2013ApJ...764....3R,
       author = {{Robertson}, Paul and {Endl}, Michael and {Cochran}, William D. and {Dodson-Robinson}, Sarah E.},
        title = "{H{\ensuremath{\alpha}} Activity of Old M Dwarfs: Stellar Cycles and Mean Activity Levels for 93 Low-mass Stars in the Solar Neighborhood}",
      journal = {\apj},
         year = 2013,
        month = feb,
       volume = {764},
       number = {1},
          eid = {3},
        pages = {3},
          doi = {10.1088/0004-637X/764/1/3},
archivePrefix = {arXiv},
       eprint = {1211.6091},
 primaryClass = {astro-ph.SR},
       adsurl = {https://ui.adsabs.harvard.edu/abs/2013ApJ...764....3R}
}

@ARTICLE{2012ApJS..203...29G,
       author = {{Godon}, Patrick and {Sion}, Edward M. and {Levay}, Karen and {Linnell}, Albert P. and {Szkody}, Paula and {Barrett}, Paul E. and {Hubeny}, Ivan and {Blair}, William P.},
        title = "{An Online Catalog of Cataclysmic Variable Spectra from the Far-Ultraviolet Spectroscopic Explorer}",
      journal = {\apjs},
         year = 2012,
        month = dec,
       volume = {203},
       number = {2},
          eid = {29},
        pages = {29},
          doi = {10.1088/0067-0049/203/2/29},
archivePrefix = {arXiv},
       eprint = {1210.1118},
 primaryClass = {astro-ph.SR},
       adsurl = {https://ui.adsabs.harvard.edu/abs/2012ApJS..203...29G}
}

@ARTICLE{2012ARep...56..716S,
       author = {{Savanov}, I.~S.},
        title = "{Activity cycles of M dwarfs}",
      journal = {Astronomy Reports},
         year = 2012,
        month = sep,
       volume = {56},
       number = {9},
        pages = {716-721},
          doi = {10.1134/S1063772912090077},
       adsurl = {https://ui.adsabs.harvard.edu/abs/2012ARep...56..716S}
}

@ARTICLE{2011AN....332..866M,
       author = {{Morgenthaler}, A. and {Petit}, P. and {Morin}, J. and {Auri{\`e}re}, M. and {Dintrans}, B. and {Konstantinova-Antova}, R. and {Marsden}, S.},
        title = "{Direct observation of magnetic cycles in Sun-like stars}",
      journal = {Astronomische Nachrichten},
         year = 2011,
        month = dec,
       volume = {332},
        pages = {866},
          doi = {10.1002/asna.201111592},
archivePrefix = {arXiv},
       eprint = {1109.3982},
 primaryClass = {astro-ph.SR},
       adsurl = {https://ui.adsabs.harvard.edu/abs/2011AN....332..866M}
}

@ARTICLE{2009A&A...501..703O,
       author = {{Ol{\'a}h}, K. and {Koll{\'a}th}, Z. and {Granzer}, T. and {Strassmeier}, K.~G. and {Lanza}, A.~F. and {J{\"a}rvinen}, S. and {Korhonen}, H. and {Baliunas}, S.~L. and {Soon}, W. and {Messina}, S. and {Cutispoto}, G.},
        title = "{Multiple and changing cycles of active stars. II. Results}",
      journal = {\aap},
         year = 2009,
        month = jul,
       volume = {501},
       number = {2},
        pages = {703-713},
          doi = {10.1051/0004-6361/200811304},
archivePrefix = {arXiv},
       eprint = {0904.1747},
 primaryClass = {astro-ph.SR},
       adsurl = {https://ui.adsabs.harvard.edu/abs/2009A&A...501..703O}
}

@ARTICLE{2008A&A...480..481B,
       author = {{Borges}, B.~W. and {Baptista}, R. and {Papadimitriou}, C. and {Giannakis}, O.},
        title = "{Cyclical period changes in HT Cassiopeiae: a difference between systems above and below the period gap}",
      journal = {\aap},
         year = 2008,
        month = mar,
       volume = {480},
       number = {2},
        pages = {481-487},
          doi = {10.1051/0004-6361:20078596},
archivePrefix = {arXiv},
       eprint = {0711.3660},
 primaryClass = {astro-ph},
       adsurl = {https://ui.adsabs.harvard.edu/abs/2008A&A...480..481B}
}

@ARTICLE{2007ApJ...657..486B,
       author = {{B{\"o}hm-Vitense}, Erika},
        title = "{Chromospheric Activity in G and K Main-Sequence Stars, and What It Tells Us about Stellar Dynamos}",
      journal = {\apj},
         year = 2007,
        month = mar,
       volume = {657},
       number = {1},
        pages = {486-493},
          doi = {10.1086/510482},
       adsurl = {https://ui.adsabs.harvard.edu/abs/2007ApJ...657..486B}
}

@ARTICLE{2003MNRAS.345..889B,
       author = {{Baptista}, R. and {Borges}, B.~W. and {Bond}, H.~E. and {Jablonski}, F. and {Steiner}, J.~E. and {Grauer}, A.~D.},
        title = "{Cyclical period changes in the dwarf novae V2051 Oph and V4140 Sgr}",
      journal = {\mnras},
         year = 2003,
        month = nov,
       volume = {345},
       number = {3},
        pages = {889-896},
          doi = {10.1046/j.1365-8711.2003.07014.x},
archivePrefix = {arXiv},
       eprint = {astro-ph/0307514},
 primaryClass = {astro-ph},
       adsurl = {https://ui.adsabs.harvard.edu/abs/2003MNRAS.345..889B}
}

@ARTICLE{2003A&A...404..301R,
       author = {{Ritter}, H. and {Kolb}, U.},
        title = "{Catalogue of cataclysmic binaries, low-mass X-ray binaries   and related objects (Seventh edition)}",
      journal = {\aap},
         year = 2003,
        month = jun,
       volume = {404},
        pages = {301-303},
          doi = {10.1051/0004-6361:20030330},
archivePrefix = {arXiv},
       eprint = {astro-ph/0301444},
 primaryClass = {astro-ph},
       adsurl = {https://ui.adsabs.harvard.edu/abs/2003A&A...404..301R}
}

@ARTICLE{2002MNRAS.335L..75B,
       author = {{Baptista}, R. and {Jablonski}, F. and {Oliveira}, E. and {Vrielmann}, S. and {Woudt}, P.~A. and {Catal{\'a}n}, M.~S.},
        title = "{Cyclical period changes in Z Chamaeleontis}",
      journal = {\mnras},
         year = 2002,
        month = sep,
       volume = {335},
       number = {3},
        pages = {L75-L78},
          doi = {10.1046/j.1365-8711.2002.05880.x},
archivePrefix = {arXiv},
       eprint = {astro-ph/0207438},
 primaryClass = {astro-ph},
       adsurl = {https://ui.adsabs.harvard.edu/abs/2002MNRAS.335L..75B}
}

@ARTICLE{2001A&A...369..882A,
       author = {{Ak}, T. and {Ozkan}, M.~T. and {Mattei}, J.~A.},
        title = "{Solar-type cycles of the secondary stars in cataclysmic variables}",
      journal = {\aap},
         year = 2001,
        month = apr,
       volume = {369},
        pages = {882-888},
          doi = {10.1051/0004-6361:20010167},
       adsurl = {https://ui.adsabs.harvard.edu/abs/2001A&A...369..882A}
}

@ARTICLE{2024ApJ...976..243D,
       author = {{Ding}, Yuedan and {Zhang}, Shidi and {Han}, Henggeng and {Cui}, Wenyuan and {Wang}, Song and {Fang}, Min and {Gao}, Yawei},
        title = "{Double-edged Sword: The Influence of Tidal Interaction on Stellar Activity in Binaries}",
      journal = {\apj},
         year = 2024,
        month = dec,
       volume = {976},
       number = {2},
          eid = {243},
        pages = {243},
          doi = {10.3847/1538-4357/ad8eb9},
archivePrefix = {arXiv},
       eprint = {2410.15039},
 primaryClass = {astro-ph.SR},
       adsurl = {https://ui.adsabs.harvard.edu/abs/2024ApJ...976..243D}
}

@ARTICLE{1995ApJ...438..269B,
       author = {{Baliunas}, S.~L. and {Donahue}, R.~A. and {Soon}, W.~H. and {Horne}, J.~H. and {Frazer}, J. and {Woodard-Eklund}, L. and {Bradford}, M. and {Rao}, L.~M. and {Wilson}, O.~C. and {Zhang}, Q. and {Bennett}, W. and {Briggs}, J. and {Carroll}, S.~M. and {Duncan}, D.~K. and {Figueroa}, D. and {Lanning}, H.~H. and {Misch}, T. and {Mueller}, J. and {Noyes}, R.~W. and {Poppe}, D. and {Porter}, A.~C. and {Robinson}, C.~R. and {Russell}, J. and {Shelton}, J.~C. and {Soyumer}, T. and {Vaughan}, A.~H. and {Whitney}, J.~H.},
        title = "{Chromospheric Variations in Main-Sequence Stars. II.}",
      journal = {\apj},
         year = 1995,
        month = jan,
       volume = {438},
        pages = {269},
          doi = {10.1086/175072},
       adsurl = {https://ui.adsabs.harvard.edu/abs/1995ApJ...438..269B}
}

@ARTICLE{1994PASP..106.1075R,
       author = {{Richman}, Hayley R. and {Applegate}, James H. and {Patterson}, Joseph},
        title = "{Long-Term Periods in Cataclysmic Variables}",
      journal = {\pasp},
         year = 1994,
        month = oct,
       volume = {106},
        pages = {1075},
          doi = {10.1086/133481},
       adsurl = {https://ui.adsabs.harvard.edu/abs/1994PASP..106.1075R}
}

@ARTICLE{1993ApJ...414L..33S,
       author = {{Soon}, W.~H. and {Baliunas}, S.~L. and {Zhang}, Q.},
        title = "{An Interpretation of Cycle Periods of Stellar Chromospheric Activity}",
      journal = {\apjl},
         year = 1993,
        month = sep,
       volume = {414},
        pages = {L33},
          doi = {10.1086/186989},
       adsurl = {https://ui.adsabs.harvard.edu/abs/1993ApJ...414L..33S}
}

@ARTICLE{1955ApJ...122..293P,
       author = {{Parker}, Eugene N.},
        title = "{Hydromagnetic Dynamo Models.}",
      journal = {\apj},
         year = 1955,
        month = sep,
       volume = {122},
        pages = {293},
          doi = {10.1086/146087},
       adsurl = {https://ui.adsabs.harvard.edu/abs/1955ApJ...122..293P}
}

@INPROCEEDINGS{1992ASPC...29..284B,
       author = {{Bianchini}, A.},
        title = "{Long-term Quasiperiodic Variability in CV's and its Possible Connection with Solar-type Cycles of Secondary Components}",
    booktitle = {Cataclysmic Variable Stars},
         year = 1992,
       editor = {{Vogt}, Nikolaus},
       series = {Astronomical Society of the Pacific Conference Series},
       volume = {29},
        month = jan,
        pages = {284},
       adsurl = {https://ui.adsabs.harvard.edu/abs/1992ASPC...29..284B}
}

@ARTICLE{1990AJ.....99.1941B,
       author = {{Bianchini}, A.},
        title = "{Solar-Type Cycles in Close Binary Systems}",
      journal = {\aj},
         year = 1990,
        month = jun,
       volume = {99},
        pages = {1941},
          doi = {10.1086/115476},
       adsurl = {https://ui.adsabs.harvard.edu/abs/1990AJ.....99.1941B}
}

@ARTICLE{1988Natur.336..129W,
       author = {{Warner}, Brian},
        title = "{Quasiperiodicity in cataclysmic variable stars caused by solar-type magnetic cycles}",
      journal = {\nat},
         year = 1988,
        month = nov,
       volume = {336},
       number = {6195},
        pages = {129-134},
          doi = {10.1038/336129a0},
       adsurl = {https://ui.adsabs.harvard.edu/abs/1988Natur.336..129W}
}

@ARTICLE{1987ApJ...322L..99A,
       author = {{Applegate}, James H. and {Patterson}, Joseph},
        title = "{Magnetic Activity, Tides, and Orbital Period Changes in Close Binaries}",
      journal = {\apjl},
         year = 1987,
        month = nov,
       volume = {322},
        pages = {L99},
          doi = {10.1086/185044},
       adsurl = {https://ui.adsabs.harvard.edu/abs/1987ApJ...322L..99A}
}

@ARTICLE{1984ApJ...287..769N,
       author = {{Noyes}, R.~W. and {Weiss}, N.~O. and {Vaughan}, A.~H.},
        title = "{The relation between stellar rotation rate and activity cycle periods.}",
      journal = {\apj},
         year = 1984,
        month = dec,
       volume = {287},
        pages = {769-773},
          doi = {10.1086/162735},
       adsurl = {https://ui.adsabs.harvard.edu/abs/1984ApJ...287..769N}
}

@ARTICLE{2016A&A...590A.133O,
       author = {{Ol{\'a}h}, K. and {K{\H{o}}v{\'a}ri}, Zs. and {Petrovay}, K. and {Soon}, W. and {Baliunas}, S. and {Koll{\'a}th}, Z. and {Vida}, K.},
        title = "{Magnetic cycles at different ages of stars}",
      journal = {\aap},
         year = 2016,
        month = jun,
       volume = {590},
          eid = {A133},
        pages = {A133},
          doi = {10.1051/0004-6361/201628479},
archivePrefix = {arXiv},
       eprint = {1604.06701},
 primaryClass = {astro-ph.SR},
       adsurl = {https://ui.adsabs.harvard.edu/abs/2016A&A...590A.133O}
}

@ARTICLE{2021arXiv210804778P,
       author = {{Paegert}, Martin and {Stassun}, Keivan G. and {Collins}, Karen A. and {Pepper}, Joshua and {Torres}, Guillermo and {Jenkins}, Jon and {Twicken}, Joseph D. and {Latham}, David W.},
        title = "{TESS Input Catalog versions 8.1 and 8.2: Phantoms in the 8.0 Catalog and How to Handle Them}",
      journal = {arXiv e-prints},
         year = 2021,
        month = aug,
          eid = {arXiv:2108.04778},
        pages = {arXiv:2108.04778},
          doi = {10.48550/arXiv.2108.04778},
archivePrefix = {arXiv},
       eprint = {2108.04778},
 primaryClass = {astro-ph.EP},
       adsurl = {https://ui.adsabs.harvard.edu/abs/2021arXiv210804778P}
}

@ARTICLE{1999ApJ...524..295S,
       author = {{Saar}, Steven H. and {Brandenburg}, Axel},
        title = "{Time Evolution of the Magnetic Activity Cycle Period. II. Results for an Expanded Stellar Sample}",
      journal = {\apj},
         year = 1999,
        month = oct,
       volume = {524},
       number = {1},
        pages = {295-310},
          doi = {10.1086/307794},
       adsurl = {https://ui.adsabs.harvard.edu/abs/1999ApJ...524..295S}
}

@ARTICLE{2003PASP..115.1105S,
       author = {{Shafter}, Allen W. and {Holland}, Julia N.},
        title = "{A Multicolor Photometric Study of the Deeply Eclipsing Dwarf Nova EX Draconis}",
      journal = {\pasp},
         year = 2003,
        month = sep,
       volume = {115},
       number = {811},
        pages = {1105-1117},
          doi = {10.1086/376934},
       adsurl = {https://ui.adsabs.harvard.edu/abs/2003PASP..115.1105S}
}

@ARTICLE{2012A&A...539A.153P,
       author = {{Pilar{\v{c}}{\'\i}k}, L. and {Wolf}, M. and {Dubovsk{\'y}}, P.~A. and {Hornoch}, K. and {Kotkov{\'a}}, L.},
        title = "{Period changes of the long-period cataclysmic binary EX Draconis}",
      journal = {\aap},
         year = 2012,
        month = mar,
       volume = {539},
          eid = {A153},
        pages = {A153},
          doi = {10.1051/0004-6361/201117972},
       adsurl = {https://ui.adsabs.harvard.edu/abs/2012A&A...539A.153P}
}

@ARTICLE{2006MNRAS.372.1129G,
       author = {{Greenhill}, J.~G. and {Hill}, K.~M. and {Dieters}, S. and {Fienberg}, K. and {Howlett}, M. and {Meijers}, A. and {Munro}, A. and {Senkbeil}, C.},
        title = "{Decrease in the orbital period of dwarf nova OY Carinae}",
      journal = {\mnras},
         year = 2006,
        month = nov,
       volume = {372},
       number = {3},
        pages = {1129-1132},
          doi = {10.1111/j.1365-2966.2006.10920.x},
archivePrefix = {arXiv},
       eprint = {astro-ph/0602331},
 primaryClass = {astro-ph},
       adsurl = {https://ui.adsabs.harvard.edu/abs/2006MNRAS.372.1129G}
}

@ARTICLE{2024ApJ...963..160Z,
       author = {{Zhao}, Xinlin and {Wang}, Song and {Li}, Xue and {Xiang}, Yue and {Xu}, Fukun and {Gu}, Shenghong and {Du}, Bing and {Liu}, Jifeng},
        title = "{Stellar Cycle and Evolution of Polar Spots in an M+WD Binary}",
      journal = {\apj},
         year = 2024,
        month = mar,
       volume = {963},
       number = {2},
          eid = {160},
        pages = {160},
          doi = {10.3847/1538-4357/ad1e64},
archivePrefix = {arXiv},
       eprint = {2401.06991},
 primaryClass = {astro-ph.SR},
       adsurl = {https://ui.adsabs.harvard.edu/abs/2024ApJ...963..160Z}
}

@ARTICLE{2003A&A...405..291H,
       author = {{Holzwarth}, V. and {Sch{\"u}ssler}, M.},
        title = "{Dynamics of magnetic flux tubes in close binary stars. I. Equilibrium and stability properties}",
      journal = {\aap},
         year = 2003,
        month = jul,
       volume = {405},
        pages = {291-301},
          doi = {10.1051/0004-6361:20030582},
archivePrefix = {arXiv},
       eprint = {astro-ph/0304496},
 primaryClass = {astro-ph},
       adsurl = {https://ui.adsabs.harvard.edu/abs/2003A&A...405..291H}
}

@ARTICLE{2003A&A...405..303H,
       author = {{Holzwarth}, V. and {Sch{\"u}ssler}, M.},
        title = "{Dynamics of magnetic flux tubes in close binary stars. II. Nonlinear evolution and surface distributions}",
      journal = {\aap},
         year = 2003,
        month = jul,
       volume = {405},
        pages = {303-311},
          doi = {10.1051/0004-6361:20030584},
archivePrefix = {arXiv},
       eprint = {astro-ph/0304498},
 primaryClass = {astro-ph},
       adsurl = {https://ui.adsabs.harvard.edu/abs/2003A&A...405..303H}
}

@ARTICLE{1996ApJ...460..848B,
       author = {{Baliunas}, S.~L. and {Nesme-Ribes}, E. and {Sokoloff}, D. and {Soon}, W.~H.},
        title = "{A Dynamo Interpretation of Stellar Activity Cycles}",
      journal = {\apj},
         year = 1996,
        month = apr,
       volume = {460},
        pages = {848},
          doi = {10.1086/177014},
       adsurl = {https://ui.adsabs.harvard.edu/abs/1996ApJ...460..848B}
}

@ARTICLE{2018A&A...616A..72W,
       author = {{Warnecke}, J.},
        title = "{Dynamo cycles in global convection simulations of solar-like stars}",
      journal = {\aap},
         year = 2018,
        month = aug,
       volume = {616},
          eid = {A72},
        pages = {A72},
          doi = {10.1051/0004-6361/201732413},
archivePrefix = {arXiv},
       eprint = {1712.01248},
 primaryClass = {astro-ph.SR},
       adsurl = {https://ui.adsabs.harvard.edu/abs/2018A&A...616A..72W}
}

@ARTICLE{2005MNRAS.357L...1B,
       author = {{Barnes}, J.~R. and {Collier Cameron}, A. and {Donati}, J.-F. and {James}, D.~J. and {Marsden}, S.~C. and {Petit}, P.},
        title = "{The dependence of differential rotation on temperature and rotation}",
      journal = {\mnras},
         year = 2005,
        month = feb,
       volume = {357},
       number = {1},
        pages = {L1-L5},
          doi = {10.1111/j.1745-3933.2005.08587.x},
archivePrefix = {arXiv},
       eprint = {astro-ph/0410575},
 primaryClass = {astro-ph},
       adsurl = {https://ui.adsabs.harvard.edu/abs/2005MNRAS.357L...1B}
}

@ARTICLE{2011MNRAS.411.1059K,
       author = {{Kitchatinov}, L.~L. and {Olemskoy}, S.~V.},
        title = "{Differential rotation of main-sequence dwarfs and its dynamo efficiency}",
      journal = {\mnras},
         year = 2011,
        month = feb,
       volume = {411},
       number = {2},
        pages = {1059-1066},
          doi = {10.1111/j.1365-2966.2010.17737.x},
archivePrefix = {arXiv},
       eprint = {1009.3734},
 primaryClass = {astro-ph.SR},
       adsurl = {https://ui.adsabs.harvard.edu/abs/2011MNRAS.411.1059K}
}

@BOOK{chandrasekhar1961,
       author = {Chandrasekhar, Subrahmanyan},
        title = {Hydrodynamic and hydromagnetic stability},
         year = 1961,
       adsurl = {https://ui.adsabs.harvard.edu/abs/1961hhs..book.....C}
}

@Article{Hubbard2012,
	author    = {W. B. Hubbard},
	journal   = {The Astrophysical Journal},
	title     = {High-Precision Maclaurin-Based Models of Rotating Liquid Planets},
	year      = {2012},
	month     = {aug},
	number    = {1},
	pages     = {L15},
	volume    = {756},
	doi       = {10.1088/2041-8205/756/1/l15},
	publisher = {American Astronomical Society}
}

@Article{Hubbard2013,
	author    = {W. B. Hubbard},
	journal   = {The Astrophysical Journal},
	title     = {CONCENTRIC {Maclaurin} SPHEROID MODELS OF ROTATING LIQUID PLANETS},
	year      = {2013},
	month     = {apr},
	number    = {1},
	pages     = {43},
	volume    = {768},
	doi       = {10.1088/0004-637x/768/1/43},
	publisher = {American Astronomical Society}
}

@Article{Wahl2017,
	author    = {Sean M. Wahl and William B. Hubbard and Burkhard Militzer},
	journal   = {Icarus},
	title     = {The Concentric Maclaurin Spheroid method with tides and a rotational enhancement of Saturn's tidal response},
	year      = {2017},
	month     = {jan},
	pages     = {183--194},
	volume    = {282},
	doi       = {10.1016/j.icarus.2016.09.011},
	publisher = {Elsevier {BV}}
}

@Article{Wahl2020,
	author    = {Sean M. Wahl and Marzia Parisi and William M. Folkner and William B. Hubbard and Burkhard Militzer},
	journal   = {The Astrophysical Journal},
	title     = {Equilibrium Tidal Response of Jupiter: Detectability by the Juno Spacecraft},
	year      = {2020},
	month     = {mar},
	number    = {1},
	pages     = {42},
	volume    = {891},
	doi       = {10.3847/1538-4357/ab6cf9},
	publisher = {American Astronomical Society},
}

@Article{Hubbard2014,
	author    = {W.B. Hubbard and G. Schubert and D. Kong and K. Zhang},
	journal   = {Icarus},
	title     = {On the convergence of the theory of figures},
	year      = {2014},
	month     = {nov},
	pages     = {138--141},
	volume    = {242},
	doi       = {10.1016/j.icarus.2014.08.014},
	publisher = {Elsevier {BV}}
}

@article{kong2022,
	author = {Kong, Dali},
	doi = {10.1103/PhysRevFluids.7.074803},
	issue = {7},
	journal = {Physical Review Fluids},
	month = {Jul},
	numpages = {12},
	pages = {074803},
	publisher = {American Physical Society},
	title = {Rapidly rotating self-gravitating Boussinesq fluid: A nonspherical model of motionless stable stratification},
	volume = {7},
	year = {2022}
}

@Article{li2023,
	author    = {Li, Wenbo and Kong, Dali},
	journal   = {Physical Review Fluids},
	title     = {Rapidly rotating self-gravitating Boussinesq fluid. {III}. A previously unknown zonal oscillation at the onset of rotating convection},
	year      = {2023},
	issn      = {2469-990X},
	month     = jan,
	number    = {1},
	pages     = {L011501},
	volume    = {8},
	doi       = {10.1103/physrevfluids.8.l011501},
	publisher     = {American Physical Society}
}

@Article{li2022,
	author        = {Li, Wenbo and Kong, Dali},
	journal       = {Physical Review Fluids},
	title         = {Rapidly rotating self-gravitating Boussinesq fluid. {II}. Onset of thermal inertial convection in oblate spheroidal cavities},
	year          = {2022},
	month         = {Oct},
	pages         = {103502},
	volume        = {7},
	doi           = {10.1103/PhysRevFluids.7.103502},
	issue         = {10},
	numpages      = {21},
	publisher     = {American Physical Society}
}

@article{li2024,
	title = {Rapidly rotating self-gravitating Boussinesq fluid. {IV}. Onset of multimodal thermal convection influenced by oblate spheroidal geometry},
	author = {Li, Wenbo and Kong, Dali},
	journal = {Phys. Rev. Fluids},
	volume = {9},
	issue = {11},
	pages = {113502},
	numpages = {17},
	year = {2024},
	month = {Nov},
	publisher = {American Physical Society},
	doi = {10.1103/PhysRevFluids.9.113502}
}

@ARTICLE{1984ApJS...54..443P,
       author = {{Patterson}, J.},
        title = "{The evolution of cataclysmic and low-mass X-ray binaries.}",
      journal = {\apjs},
         year = 1984,
        month = apr,
       volume = {54},
        pages = {443-493},
          doi = {10.1086/190940},
       adsurl = {https://ui.adsabs.harvard.edu/abs/1984ApJS...54..443P}
}

@ARTICLE{2011ApJ...743...48W,
       author = {{Wright}, Nicholas J. and {Drake}, Jeremy J. and {Mamajek}, Eric E. and {Henry}, Gregory W.},
        title = "{The Stellar-activity-Rotation Relationship and the Evolution of Stellar Dynamos}",
      journal = {\apj},
         year = 2011,
        month = dec,
       volume = {743},
       number = {1},
          eid = {48},
        pages = {48},
          doi = {10.1088/0004-637X/743/1/48},
archivePrefix = {arXiv},
       eprint = {1109.4634},
 primaryClass = {astro-ph.SR},
       adsurl = {https://ui.adsabs.harvard.edu/abs/2011ApJ...743...48W}
}

@ARTICLE{2025NatAs...9.1045Y,
       author = {{Yu}, Jie and {Gehan}, Charlotte and {Hekker}, Saskia and {Bazot}, Mich{\"a}el and {Cameron}, Robert H. and {Gaulme}, Patrick and {Bedding}, Timothy R. and {Murphy}, Simon J. and {Han}, Zhanwen and {Ting}, Yuan-Sen and {Tayar}, Jamie and {Chen}, Yajie and {Gizon}, Laurent and {Nordhaus}, Jason and {Bi}, Shaolan},
        title = "{Enhanced magnetic activity in rapidly rotating binary stars}",
      journal = {Nature Astronomy},
         year = 2025,
        month = jul,
       volume = {9},
        pages = {1045-1052},
          doi = {10.1038/s41550-025-02562-2},
archivePrefix = {arXiv},
       eprint = {2505.19967},
 primaryClass = {astro-ph.SR},
       adsurl = {https://ui.adsabs.harvard.edu/abs/2025NatAs...9.1045Y}
}

@ARTICLE{2022MNRAS.514.4932C,
       author = {{Chahal}, Deepak and {de Grijs}, Richard and {Kamath}, Devika and {Chen}, Xiaodian},
        title = "{Statistics of BY Draconis chromospheric variable stars}",
      journal = {\mnras},
         year = 2022,
        month = aug,
       volume = {514},
       number = {4},
        pages = {4932-4943},
          doi = {10.1093/mnras/stac1660},
archivePrefix = {arXiv},
       eprint = {2206.05505},
 primaryClass = {astro-ph.SR},
       adsurl = {https://ui.adsabs.harvard.edu/abs/2022MNRAS.514.4932C}
}

\clearpage

\begin{appendix}

%




\renewcommand{\thesection}{Supplementary Materials}
\renewcommand{\thesubsection}{\Alph{subsection}}
\setcounter{table}{0}   
\setcounter{figure}{0}
\renewcommand{\thetable}{S\arabic{table}}
\renewcommand{\thefigure}{S\arabic{figure}}
\section{}

\subsection{Period values for our sample of M dwarfs.} 

During sample construction, we cross‑matched M single stars with SIMBAD and found six objects catalogued as spectroscopic binaries: BD-022198, GJ182, HIP1910, HIP23200, HIP36349, V1428Aql. They were moved to the wide binary sample.

Tables \ref{tableS1.tab}, \ref{tableS2.tab}, and \ref{tableS3.tab} list the period estimates for the M single dwarfs, M dwarfs in wide binaries, and M dwarfs in close binaries, respectively.

The methods for cycle period measurements listed in these tables are as follows. LC: light curve; CL: chromospheric line. LS: Lomb Scargle; GLS: Generalized Lomb Scargle; BTFT: Bilinear Time-Frequency Transformation; PDM: Phase Dispersion Minimization; STFT: Short-Term Fourier Transform; FT: Fourier Transform; DFT: Discrete Fourier Transform;  IO: intervals between outbursts.

The references for period measurements listed in these tables include: 
(1) \cite{2019A&A...621A.126D}; 
(2) \cite{2013AN....334..972V}; 
(3) \cite{2018A&A...612A..89S}; 
(4) \cite{2016A&A...595A..12S}; 
(5) \cite{2019A&A...622A..40K}; 
(6) \cite{2013ApJ...764....3R}; 
(7) \cite{2016A&A...590A.133O};
(8) \cite{2016A&A...591A..43D};
(9) \cite{2017A&A...606A..58D};
(10) \cite{2014MNRAS.441.2744V};
(11) \cite{2017MNRAS.464.3281W};
(12) \cite{1992ASPC...29..284B};
(13) \cite{1999ApJ...524..295S};
(14) \cite{2001A&A...369..882A};
(15) \cite{2003A&A...404..301R};
(16) \cite{1990AJ.....99.1941B};
(17) \cite{2012ApJS..203...29G};
(18) \cite{2003PASP..115.1105S};
(19) \cite{2012A&A...539A.153P};
(20) \cite{2022MNRAS.510.4736K};
(21) \cite{2008A&A...480..481B};
(22) \cite{2006MNRAS.372.1129G};
(23) \cite{1994PASP..106.1075R};
(24) \cite{2024ApJ...963..160Z};
(25) \cite{2003MNRAS.345..889B};
(26) \cite{2002MNRAS.335L..75B};
(27) \cite{2021arXiv210804778P};
(28) \cite{2009A&A...501..703O};
(29) \cite{2007ApJ...657..486B};
(30) \cite{2016MNRAS.463.1342S}.

For stars with multiple cycle measurements, $P_{\rm cyc1}$ represents the most recent result. For MV Lyr, RW Tri, and V603 Aql, the referenced cycle periods are 11--12 years, 8--14 years, and 15--20 years, respectively; we used the median value of each range as the cycle period and half of the interval as its uncertainty.

\begin{table*}
\caption{Period measurements for M-type single stars.
\label{tableS1.tab}}
\begin{center}
\setlength{\tabcolsep}{1.pt}
\renewcommand{\arraystretch}{1.2}
\begin{tabular}{lrrcccccccll}
\hline\noalign{\smallskip}
Name & R.A.  & Decl. & Type  & $P_{\rm rot}$ & log($\frac{1}{P_{\rm rot}}$) & $P_{\rm cyc1}$ & $P_{\rm cyc2}$ & log($\frac{P_{\rm cyc1}}{P_{\rm rot}}$) & log($\frac{P_{\rm cyc2}}{P_{\rm rot}}$) & Method$^a$ & Ref. \\
 & (deg) & (deg) & & (day) & (1/day) &   (day) & (day) &  &  &   & \\
 \hline\noalign{\smallskip}        
ANSex & 153.07362 & $-$3.74566 & M2 & 21.6 & $-$1.33 & 1168 & --- & 1.73 & --- & LC(LS) &(1)\\
BD+044157 & 293.666 & 4.58251 & M0 & 12.9 &  $-$1.11 & 1095 & --- & 1.93 & --- & LC(LS) &(1)\\
EYDra & 274.06991 & 54.1727 & M1 & 0.46 & 0.34 & 348 & --- & 2.88 & --- & LC(FT) &(2)\\
FIVir & 176.93499 & 0.80456 & M4 & 163.0 & $-$2.21 & 1642 & --- & 1.0 & --- & LC(LS) &(1)\\
GJ162 & 62.15583 & 33.63704 & M1 & 32.4 &  $-$1.51 & 1752 & --- & 1.73 & --- & CL(GLS) &(3) \\
GJ176 & 70.7324 & 18.95817 & M2 & 40.8 &  $-$1.61 & 2153 & --- & 1.72 & --- & LC(GLS) &(4)\\
GJ184 & 75.84959 & 53.12847 & M0 & 45.0 &  $-$1.65 & 1277 & --- & 1.45 & --- & CL(GLS) &(3)\\
GJ2 & 1.29537 & 45.78657 & M1 & 21.2 &  $-$1.33 & 1168 & --- & 1.74 & --- & CL(GLS) &(3)\\
GJ205 & 82.86415 & $-$3.67723 & M1 & 33.5 &  $-$1.52 & 1278 & --- & 1.58 & --- & LC(BTFT) &(5)\\
GJ2066 & 124.03326 & 1.30257 & M2 & 40.7 &  $-$1.61 & 1496 & --- & 1.57 & --- & LC(LS) &(1)\\
GJ229 & 92.64423 & $-$21.86463 & M1 & 27.3 &  $-$1.44 & 3066 & --- & 2.05 & --- & LC(GLS) &(4)\\
GJ270 & 109.88031 & 32.83009 & M2 & 30.0 &  $-$1.48 & 2687 & --- & 1.95 & --- & CL(GLS) &(5,6)\\
GJ273 & 111.85208 & 5.22579 & M3 & 115.9 & $-$2.06 & 2409 & --- & 1.32 & --- & LC(GLS) &(4)\\
GJ285 & 116.16738 & 3.55247 & M4 & 2.8 &  $-$0.45 & 3869 & --- & 3.14 & --- & LC(GLS) &(4)\\
GJ328 & 133.78176 & 1.5465 & M0 & 33.6 &  $-$1.53 & 2000 & --- & 1.77 & --- & CL(GLS) &(5,6)\\
GJ3367 & 86.82486 &  $-$0.01355 & M0 & 12.1 &  $-$1.08 & 2411 & --- & 2.3 & --- & LC(BTFT) &(5)\\
GJ358 & 144.9432 & $-$41.06756 & M2 & 26.0 &  $-$1.42 & 1680 & --- & 1.81 & --- & LC(BTFT) &(5)\\
GJ382 & 153.07362 & $-$3.74566 & M2 & 21.2 &  $-$1.33 & 4964 & --- & 2.37 & --- & LC(GLS) &(4)\\
GJ3822 & 210.58177 & 13.68964 & M0 & 18.3 &  $-$1.26 & 4635 & --- & 2.4 & --- & CL(GLS) &(3)\\
GJ3997 & 258.95874 & 19.00004 & M0 & 37.0 &  $-$1.57 & 803 & --- & 1.34 & --- & CL(GLS) &(3)\\
GJ3998 & 259.00265 & 11.05767 & M1 & 33.6 &  $-$1.53 & 657 & --- & 1.29 & --- & CL(GLS) &(3)\\
GJ4057 & 276.26998 & 24.63458 & M0 & 26.7 &  $-$1.43 & 949 & --- & 1.55 & --- & CL(GLS) &(3)\\
GJ4306 & 343.99936 & 17.81105 & M1 & 27.0 &  $-$1.43 & 1606 & --- & 1.77 & --- & CL(GLS) &(3)\\
GJ431 & 172.9438 & $-$41.04644 & M3 & 14.3 &  $-$1.16 & 1826 & --- & 2.11 & --- & LC(BTFT) &(5)\\
GJ447 & 176.93499 & 0.80456 & M4 & 165.1 & $-$2.22 & 1496 & --- & 0.96 & --- & LC(GLS) &(4)\\
GJ479 & 189.46757 & $-$52.00148 & M3 & 22.5 &  $-$1.35 & 730 & --- & 1.51 & --- & LC(GLS) &(4)\\
GJ514 & 202.49911 & 10.37716 & M1 & 28.0 &  $-$1.45 & 3613 & --- & 2.11 & --- & LC(GLS) &(4)\\
GJ526 & 206.4324 & 14.89152 & M4 & 52.3 &  $-$1.72 & 3613 & --- & 1.84 & --- & LC(GLS) &(4)\\
GJ551 & 217.42894 & $-$62.67949 & M5 & 83.2 &  $-$1.92 & 2482 & --- & 1.47 & --- & LC(GLS) &(4)\\
GJ581 & 229.86178 & $-$7.72227 & M3 & 130.0 & $-$2.11 & 1633 & --- & 1.1 & --- & CL(GLS) &(5,6)\\
\noalign{\smallskip}\hline
  \end{tabular}
  \end{center}
$^a$NOTES. LC: light curve; CL: chromospheric line. LS: Lomb Scargle; GLS: Generalized Lomb Scargle; BTFT: Bilinear Time-Frequency Transformation; PDM: Phase Dispersion Minimization; STFT: Short-Term Fourier Transform; FT: Fourier Transform; DFT: Discrete Fourier Transform;  IO: intervals between outbursts.
\end{table*}

\setcounter{table}{0}
 \begin{table*}
 \caption{Continued. Period measurements for M-type single stars.}
 \centering
\setlength{\tabcolsep}{0.8pt}
\renewcommand{\arraystretch}{1.2}
  \begin{tabular}{lcccccccccll}
 \hline\noalign{\smallskip}
Name & R.A.  & Decl. & Type  & $P_{\rm rot}$ & log($\frac{1}{P_{\rm rot}}$) & $P_{\rm cyc1}$ & $P_{\rm cyc2}$ & log($\frac{P_{\rm cyc1}}{P_{\rm rot}}$) & log($\frac{P_{\rm cyc2}}{P_{\rm rot}}$) & Method & Refs. \\
 & (deg) & (deg) & & (day) & (1/day) &   (day) & (day) &  &  &   & \\
 \hline\noalign{\smallskip}     
GJ588 & 233.05388 & $-$41.27559 & M2 & 61.3 &  $-$1.79 & 1899 & --- & 1.49 & --- & LC(GLS) &(4)\\
GJ625 & 246.3526 & 54.3041 & M2 & 77.8 &  $-$1.89 & 1204 & --- & 1.19 & --- & CL(GLS) &(3)\\
GJ628 & 247.57524 &  $-$12.66259 & M4 & 119.3 & $-$2.08 & 1606 & --- & 1.13 & --- & LC(GLS) &(4)\\
 GJ674 & 262.16644 & $-$46.89519 & M3 & 35.0 &  $-$1.54 & 3686 & --- & 2.02 & --- & LC(GLS) &(4)\\
GJ729 & 282.45568 & $-$23.83624 & M3 & 2.87 &  $-$0.46 & 1790 & --- & 2.8 & --- & LC(BTFT) &(5)\\
GJ740 & 284.50057 & 5.90812 & M0 & 36.3 &  $-$1.56 & 2044 & --- & 1.75 & --- & CL(GLS) &(3)\\
GJ803 & 311.28972 & $-$31.3409 & M1 & 4.86 &  $-$0.69 & 1753 & --- & 2.56 & --- & LC(BTFT) &(5)\\
GJ832 & 323.39156 & $-$49.009 & M1 & 45.7 &  $-$1.66 & 4818 & --- & 2.02 & --- & LC(GLS) &(4)\\
GJ846 & 330.54281 & 1.40023 & M0 & 31.0 &  $-$1.49 & 3102 & --- & 2.0 & --- & LC(GLS) &(4)\\
GJ849 & 332.4181 & $-$4.64074 & M3 & 39.2 &  $-$1.59 & 3723 & --- & 1.98 & --- & LC(GLS) &(4)\\
GJ890 & 347.08146 &  $-$15.40994 & M0 & 0.43 & 0.37 & 1424 & --- & 3.52 & --- & LC(BTFT) &(5)\\
HD197481 & 311.28972 & $-$31.3409 & M1 & 4.9 &  $-$0.69 & 2774 & --- & 2.75 & --- & LC(GLS) &(4)\\
HD95735 & 165.83414 & 35.96988 & M2 & 54.0 &  $-$1.73 & 1424 & --- & 1.42 & --- & LC(LS) &(7)\\
HIP102409 & 311.28972 & $-$31.3409 & M1 & 4.84 &  $-$0.68 & 1428 & --- & 2.47 & --- & LC(LS/PDM) &(8,9)\\
HIP107345 & 326.12551 & $-$60.97747 & M0 & 4.55 &  $-$0.66 & 1600 & --- & 2.55 & --- & LC(LS/PDM) &(8,9)\\
HIP17695 & 56.84725 &  $-$1.97221 & M3 & 3.88 &  $-$0.59 & 1826 & --- & 2.67 & --- & LC(BTFT) &(5)\\
HIP23309 & 75.19637 & $-$57.25707 & M0 & 8.74 &  $-$0.94 & 1666 & --- & 2.28 & --- & LC(LS/PDM) &(8,9)\\
ILAqr & 343.31972 &  $-$14.2637 & M3 & 81.0 &  $-$1.91 & 1642 & --- & 1.31 & --- & LC(LS) &(1)\\
KIC03541346 & 290.48612 & 38.62535 & M0 & 0.91 & 0.04 & 330 & --- & 2.56 & --- & LC(STFT) &(10)\\
KIC04819564 & 286.35546 & 39.91813 & M0 & 0.38 & 0.42 & 530 & --- & 3.14 & --- & LC(STFT) &(10)\\
KIC04953358 & 298.77777 & 40.09237 & M1 & 0.65 & 0.19 & 600 & --- & 2.97 & --- & LC(STFT) &(10)\\
KIC05791720 & 292.07429 & 41.07007 & M3 & 0.77 & 0.12 & 320 & --- & 2.62 & --- & LC(STFT) &(10)\\
KIC06675318 & 286.34714 & 42.11492 & M0 & 0.58 & 0.24 & 370 & --- & 2.81 & --- & LC(STFT) &(10)\\
KIC07592990 & 286.36267 & 43.24619 & M0 & 0.44 & 0.35 & 500 & --- & 3.05 & --- & LC(STFT) &(10)\\
KIC08314902 & 296.89699 & 44.23397 & M0 & 0.81 & 0.09 & 330 & --- & 2.61 & --- & LC(STFT) &(10)\\
KIC10515986 & 282.33037 & 47.72497 & M1 & 0.75 & 0.13 & 350 & --- & 2.67 & --- & LC(STFT) &(10)\\
KIC11087527 & 293.5498 & 48.68472 & M0 & 0.41 & 0.39 & 310 & --- & 2.88 & --- & LC(STFT) &(10)\\
LP 816-60 & 313.13757 &  $-$16.97472 & M4 & 67.6 &  $-$1.83 & 3869 & --- & 1.76 & --- & LC(GLS) &(4)\\
OTSer & 230.47055 & 20.97775 & M1 & 3.37 &  $-$0.53 & 2372 & --- & 2.85 & --- & LC(LS) &(1)\\
ProxCen & 217.42894 & $-$62.67949 & M3 & 83.0 &  $-$1.92 & 2557 & --- & 1.49 & --- & LC(LS) &(11)\\
TWA2 & 167.30747 & $-$30.02777 & M2 & 4.85 &  $-$0.69 & 1489 & --- & 2.49 & --- & LC(LS/PDM) &(8,9)\\
TWA25 & 183.878 & $-$39.81183 & M1 & 5.06 &  $-$0.7 & 1666 & --- & 2.52 & --- & LC(LS/PDM) &(8,9)\\
TYC5832-0666-1 & 353.12859 &  $-$12.2643 & M0 & 5.69 &  $-$0.76 & 1695 & --- & 2.47 & --- & LC(LS/PDM) &(8,9)\\
TYC9073-0762-1 & 281.71897 & $-$62.17684 & M1 & 5.37 &  $-$0.73 & 1999 & --- & 2.57 & --- & LC(LS/PDM) &(8,9)\\
V2306Oph & 247.57524 &  $-$12.66259 & M3 & 119.0 & $-$2.08 & 1423 & --- & 1.08 & --- & LC(LS) &(1)\\
\noalign{\smallskip}\hline
  \end{tabular}
\end{table*}

\begin{table*}
\caption{Period measurements for wide binaries including M-type stars.
\label{tableS2.tab}}
\begin{center}
\setlength{\tabcolsep}{1.2pt}
\renewcommand{\arraystretch}{1.2}
\begin{tabular}{lcccccccccll}
\hline\noalign{\smallskip}
Name & R.A.  & Decl. & Type  & $P_{\rm rot}$ & log($\frac{1}{P_{\rm rot}}$) & $P_{\rm cyc1}$ & $P_{\rm cyc2}$ & log($\frac{P_{\rm cyc1}}{P_{\rm rot}}$) & log($\frac{P_{\rm cyc2}}{P_{\rm rot}}$) & Method & Refs. \\
 & (deg) & (deg) & & (day) & (1/day) &   (day) & (day) &  &  &   & \\
 \hline\noalign{\smallskip}          
BD-022198 & 114.02947 & $-$3.11076 & M1 & 12.2 &  $-$1.09 & 4197 & --- & 2.54 & --- & LC(LS) &(1)\\
GJ1054A & 46.9823 & $-$28.21971 & M0 & 7.4 &  $-$0.87 & 1278 & --- & 2.24 & --- & LC(BTFT) &(5)\\
GJ1264A & 327.27399 & $-$72.10253 & M0 & 6.66 &  $-$0.82 & 1644 & --- & 2.39 & --- & LC(BTFT) &(5)\\
GJ15A & 4.59535 & 44.02295 & M1 & 45.0 &  $-$1.65 & 1022 & --- & 1.36 & --- & CL(GLS) &(3)\\
GJ182 & 74.89514 & 1.78352 & M0 & 4.43 &  $-$0.65 & 1753 & --- & 2.6 & --- & LC(BTFT) &(5)\\
GJ2036A & 73.38001 & $-$55.86031 & M2 & 2.98 &  $-$0.47 & 1790 & --- & 2.78 & --- & LC(BTFT) &(5)\\
GJ234 & 97.34746 & $-$2.81356 & M4 & 8.1 &  $-$0.91 & 2153 & --- & 2.42 & --- & LC(GLS) &(4)\\
GJ3331A & 76.70798 & $-$21.58587 & M1 & 13.7 &  $-$1.14 & 1242 & --- & 1.96 & --- & LC(BTFT) &(5)\\
GJ494 & 195.19399 & 12.37574 & M0 & 2.89 &  $-$0.46 & 1278 & --- & 2.65 & --- & LC(BTFT) &(5)\\
GJ521A & 204.85043 & 46.18649 & M1 & 49.5 &  $-$1.69 & 657 & --- & 1.12 & --- & CL(GLS) &(3)\\
GJ618A & 245.01461 & $-$37.529 & M3 & 57.4 &  $-$1.76 & 1242 & --- & 1.34 & --- & LC(BTFT) &(5)\\
GJ676A & 262.54668 & $-$51.63698 & M0 & 41.2 &  $-$1.61 & 2737 & --- & 1.82 & --- & LC(GLS) &(4)\\
GJ720A & 278.82663 & 45.74404 & M0 & 34.5 &  $-$1.54 & 839 & --- & 1.39 & --- & CL(GLS) &(3)\\
GJ752A & 289.23024 & 5.1689 & M3 & 46.0 &  $-$1.66 & 3394 & --- & 1.87 & --- & LC(GLS) &(4)\\
GJ84 & 31.27033 &  $-$17.61465 & M2 & 43.9 &  $-$1.64 & 1096 & --- & 1.4 & --- & LC(BTFT) &(5)\\
GJ841A & 329.42168 & $-$51.00618 & M2 & 1.12 &  $-$0.05 & 1278 & --- & 3.06 & --- & LC(BTFT) &(5)\\
GJ867A & 339.68989 & $-$20.62114 & M1 & 4.22 &  $-$0.63 & 1461 & --- & 2.54 & --- & LC(BTFT) &(5)\\
GJ897A & 353.19575 &  $-$16.75343 & M2 & 4.83 &  $-$0.68 & 1680 & --- & 2.54 & --- & LC(BTFT) &(5)\\
HD42581A & 92.64423 & $-$21.86463 & M1 & 27.3 &  $-$1.44 & 3029 & --- & 2.05 & --- & LC(LS) &(1)\\
HIP1910 & 6.03737 & $-$62.18454 & M0 & 1.75 &  $-$0.24 & 1025 & --- & 2.77 & --- & LC(LS/PDM) &(8,9)\\
HIP23200 & 74.89514 & 1.78352 & M0 & 4.42 &  $-$0.65 & 1597 & --- & 2.56 & --- & LC(LS/PDM) &(8,9)\\
HIP36349 & 112.21398 & $-$30.24703 & M1 & 1.64 &  $-$0.21 & 1904 & --- & 3.06 & --- & LC(LS/PDM) &(8,9)\\
TWA13A & 170.32175 & $-$34.77931 & M1 & 5.44 &  $-$0.74 & 1250 & --- & 2.36 & --- & LC(LS/PDM) &(8,9)\\
V1428Aql & 289.23024 & 5.1689 & M3 & 46.0 &  $-$1.66 & 1204 & --- & 1.42 & --- & LC(LS) &(1)\\
\noalign{\smallskip}\hline
  \end{tabular}
  \end{center}
\end{table*}

\begin{table*}
\caption{Period measurements for close binaries including M-type stars.
\label{tableS3.tab}}
\begin{center}
\footnotesize
\setlength{\tabcolsep}{1.pt}
\renewcommand{\arraystretch}{1.2}
\begin{tabular}{lcccccccccll}
\hline\noalign{\smallskip}
Name & R.A.  & Decl. & Type  & $P_{\rm rot}$ & log($\frac{1}{P_{\rm rot}}$) & $P_{\rm cyc1}^a$ & $P_{\rm cyc2}$ & log($\frac{P_{\rm cyc1}}{P_{\rm rot}}$) & log($\frac{P_{\rm cyc2}}{P_{\rm rot}}$) & Method & Refs. \\
 & (deg) & (deg) & & (day) & (1/day) &   (day) & (day) &  &  &   & \\
 \hline\noalign{\smallskip}          
ARAnd & 26.26363 & 37.94257 & M5 & 0.11 & 0.97 & 3394 & --- & 4.5 & --- &LC(IO)&(12,13)\\ 
CZOri & 94.1801 & 15.40314 & M3 & 0.22 & 0.66 & 6241$\pm$328 & --- & 4.46$\pm$0.02 & --- & LC(FT) &(14,15)\\
DQHer & 271.87605 & 45.85905 & M3 & 0.19 & 0.71 & 4891 & --- & 4.4 & --- & LC(O-C) &(16,12,13,15,17)\\
EXDra & 271.05937 & 67.90341 & M2 & 0.21 & 0.68 & 7665 & 1825 & 4.56 & 3.94 & LC(O-C) &(18,19)\\
EXHya & 193.10093 & $-$29.24889 & M5 & 0.07 & 1.17 & 10475 & 6935 & 5.19 & 5.01 & LC(O-C/FT) &(16,12,13,15,17,14)\\
EXO0748-676 & 117.14046 & $-$67.75214 & M & 0.16 & 0.8 & 4380 & --- & 4.44 & --- & LC(O-C) &(13,20)\\
HTCas & 17.55467 & 60.07652 & M5 & 0.07 & 1.13 & 13140 & --- & 5.25 & --- & LC(O-C) &(15,21)\\
MVLyr & 286.81787 & 44.01885 & M5 & 0.14 & 0.86 & 4197$\pm$182 & --- & 4.48$\pm$0.02 & --- & LC(DFT) &(16,12,15,17)\\
OYCar & 151.592 & $-$70.23461 & M9 & 0.06 & 1.2 & 12775 & 2299 & 5.31 & 4.56 & LC(FT) &(14,15,22)\\
RRPic & 98.90026 & $-$62.64008 & M4 & 0.14 & 0.84 & 5110 & --- & 4.55 & --- & LC(DFT) &(16,12,23,13,15)\\
RWTri & 36.40065 & 28.09747 & M0 & 0.23 & 0.63 & 4015$\pm$1095 & --- & 4.24$\pm$0.12 & --- & LC(O-C) &(16,12,15)\\
SSAur & 93.34347 & 47.74037 & M1 & 0.18 & 0.74 & 3869$\pm$219 & --- & 4.33$\pm$0.02 & --- &LC(IO)&(16,12,14,15,17)\\
SUUMa & 123.11779 & 62.60623 & M5 & 0.08 & 1.12 & 2190 & --- & 4.46 & --- &LC(IO)&(12,13)\\
TIC16320250 & 231.95199 & 35.61592 & M0 & 0.26 & 0.59 & 3650$\pm$365 & --- & 4.15$\pm$0.04 & --- & LC &(24)\\
TTCrt & 173.69657 &  $-$11.75844 & M0 & 0.27 & 0.57 & 1569$\pm$328 & --- & 3.77$\pm$0.09 & --- & LC(FT) &(14,15)\\
UGem & 118.7718 & 22.0014 & M5 & 0.18 & 0.75 & 2043$\pm$255 & --- & 4.06$\pm$0.05 & --- & LC(FT) &(14,15)\\
UXUMa & 204.17064 & 51.91373 & M1 & 0.2 & 0.71 & 10950 & 2591 & 4.75 & 4.12 & LC(O-C) &(16,12,23,13,15)\\
V2051Oph & 257.07953 & $-$25.80881 & M8 & 0.06 & 1.2 & 8030 & --- & 5.11 & --- & LC(O-C) &(15,25)\\
V436Cen & 168.50024 & $-$37.67993 & M9 & 0.06 & 1.2 & 9599$\pm$292 & --- & 5.19$\pm$0.01 & --- & LC(FT) &(14,15)\\
V603Aql & 282.22765 & 0.58413 & M4 & 0.14 & 0.84 & 6387$\pm$912 & --- & 4.64$\pm$0.06 & --- & LC &(23,13,15)\\
VWHyi & 62.29749 & $-$71.29488 & M5 & 0.07 & 1.13 & 3613 & --- & 4.69 & --- &LC(IO)&(12,13,15,17)\\
WWCet & 2.85322 &  $-$11.47867 & M5 & 0.18 & 0.76 & 2043$\pm$109 & --- & 4.07$\pm$0.02 & --- & LC(FT) &(14,15)\\
ZCha & 121.86563 & $-$76.53352 & M5 & 0.07 & 1.13 & 10220 & --- & 5.14 & --- & LC(FT) &(12,23,15,14,26)\\
\noalign{\smallskip}\hline
  \end{tabular}
  \end{center}
  {$^a$ For stars with multiple cycle measurements, $P_{\rm cyc1}$ represents the most recent result. For MV Lyr, RW Tri, and V603 Aql, the referenced cycle periods are 11–12 years, 8–14 years, and 15–20 years, respectively; we used the median value of each range as the cycle period and half of the interval as its uncertainty.}
\end{table*}

\subsection{Derived parameters for our binary sample including M dwarfs.}

Tables \ref{tableS4.tab} and \ref{tableS5.tab} list derived parameters for wide and close binaries with M-dwarf components, including stellar types, component masses, orbital periods, etc.

\begin{table*}
\caption{Derived parameters for wide binaries including M-type stars.
\label{tableS4.tab}}
\centering
\setlength{\tabcolsep}{5pt}
\renewcommand{\arraystretch}{1.2}
\begin{tabular}{lccccccccl}
\hline\noalign{\smallskip}
Name & Type$_{M}$  & $M$ & $R$ & Type$_{m_s}$  & $m_s$ &  $P_{\rm orb}$  &  $q_{\rm{tid}}$  &  $F_{r}$  & Refs. \\
 &  & (M$_{\odot}$) & (R$_{\odot}$)  &  &  (M$_{\odot}$) & (day) &  &  &   \\
 \hline\noalign{\smallskip}          
GJ1054A & M0 & 0.61 & 0.64 & M3 & 0.28 & 2.11 &  $-$1.23e$-$03 & 1.29e$-$03 &(5, 27)\\
GJ234 & M4 & 0.22 & 0.26 & M7 & 0.11 & 6058.0 & $-$2.81e$-$11 & 2.85e$-$11 &(4, 27)\\
GJ494 & M0 & 0.56 & 0.59 & M7 & 0.09 & 5004.0 & $-$8.17e$-$11 & 1.97e$-$10 &(5, 27)\\
GJ84 & M3 & 0.46 & 0.5 & M7 & 0.11 & 6818.0 & $-$4.49e$-$11 & 7.81e$-$11 &(5, 27)\\
GJ841A & M3 & 0.6 & 0.63 & WD & 0.6 & 1.12 & $-$6.62e$-$03 & 4.40e$-$03 &(5, 27)\\
GJ867A & M0 & 0.59 & 0.62 & M3 & 0.32 & 4.08 & $-$3.47e$-$04 & 3.27e$-$04 &(5, 27)\\
\noalign{\smallskip}\hline
  \end{tabular}
\end{table*}

\begin{table*}
\caption{Derived parameters for close binaries including M-type stars.
\label{tableS5.tab}}%
\centering
\setlength{\tabcolsep}{3pt}
\renewcommand{\arraystretch}{1.2}
\begin{tabular}{lccccccccl}
\hline\noalign{\smallskip}
Name & Type$_{M}$  & $M$ & $R$ & Type$_{m_s}$  & $m_s$ &  $P_{\rm orb}$  &  $q_{\rm{tid}}$  &  $F_{r}$  & Refs. \\
 &  & (M$_{\odot}$) & (R$_{\odot}$)  &  &  (M$_{\odot}$) & (day) &  &  &   \\
 \hline\noalign{\smallskip}          
DQHer & M3 & 0.4 & 0.44 & WD & 0.6 & 0.19 &  $-$1.35e$-$01 & 7.52e$-$02 &(16, 12, 13, 15, 17)\\
EXDra & M2 & 0.54 & 0.57 & WD & 0.72 & 0.21 &  $-$1.84e$-$01 & 1.08e$-$01 &(18, 19)\\
EXHya & M5 & 0.11 & 0.13 & WD & 0.79 & 0.07 &  $-$1.72e$-$01 & 6.53e$-$02 &(16, 12, 13, 15, 17, 14)\\
EXO0748-676 & M & 0.44 & 0.48 & NS & 2.0 & 0.16 & $-$3.23e$-$01 & 1.31e$-$01 &(13, 20)\\
HTCas & M5 & 0.09 & 0.11 & WD & 0.61 & 0.07 &  $-$1.07e$-$01 & 4.10e$-$02 &(15, 21)\\
MVLyr & M5 & 0.3 & 0.34 & WD & 0.73 & 0.14 &  $-$1.95e$-$01 & 9.15e$-$02 &(16, 12, 15, 17)\\
OYCar & M9 & 0.09 & 0.11 & WD & 0.64 & 0.06 &  $-$1.37e$-$01 & 5.20e$-$02 &(14, 15, 22)\\
RRPic & M4 & 0.4 & 0.44 & WD & 0.95 & 0.14 & $-$2.83e$-$01 & 1.34e$-$01 &(16, 12, 23, 13, 15)\\
RWTri & M0 & 0.35 & 0.39 & WD & 0.55 & 0.23 & $-$7.66e$-$02 & 4.18e$-$02 &(16, 12, 15)\\
SSAur & M1 & 0.39 & 0.43 & WD & 1.08 & 0.18 &  $-$1.83e$-$01 & 8.30e$-$02 &(16, 12, 15, 17)\\
TIC16320250 & M0 & 0.56 & 0.59 & WD & 0.67 & 0.26 &  $-$1.25e$-$01 & 7.65e$-$02 &(24)\\
UGem & M5 & 0.42 & 0.46 & WD & 1.2 & 0.18 & $-$2.18e$-$01 & 9.80e$-$02 &(14, 15)\\
UXUMa & M1 & 0.39 & 0.43 & WD & 0.9 & 0.2 &  $-$1.46e$-$01 & 6.99e$-$02 &(16, 12, 23, 13, 15)\\
V2051Oph & M8 & 0.15 & 0.18 & WD & 0.78 & 0.06 & $-$3.44e$-$01 & 1.37e$-$01 &(15, 25)\\
V436Cen & M9 & 0.17 & 0.2 & WD & 0.7 & 0.06 & $-$4.07e$-$01 & 1.69e$-$01 &(14, 15)\\
V603Aql & M4 & 0.29 & 0.33 & WD & 1.2 & 0.14 &  $-$1.88e$-$01 & 7.77e$-$02 &(23, 13, 15)\\
VWHyi & M5 & 0.11 & 0.14 & WD & 0.67 & 0.07 &  $-$1.47e$-$01 & 5.72e$-$02 &(12, 13, 15, 17)\\
WWCet & M5 & 0.41 & 0.45 & WD & 0.83 & 0.18 &  $-$1.91e$-$01 & 9.52e$-$02 &(14, 15)\\
ZCha & M5 & 0.12 & 0.15 & WD & 0.84 & 0.07 &  $-$1.86e$-$01 & 7.13e$-$02 &(12, 23, 15, 14, 26)\\
\noalign{\smallskip}\hline
  \end{tabular}
\end{table*}

\newpage

\begin{figure*}
    \center
    \includegraphics[width=0.9\textwidth]{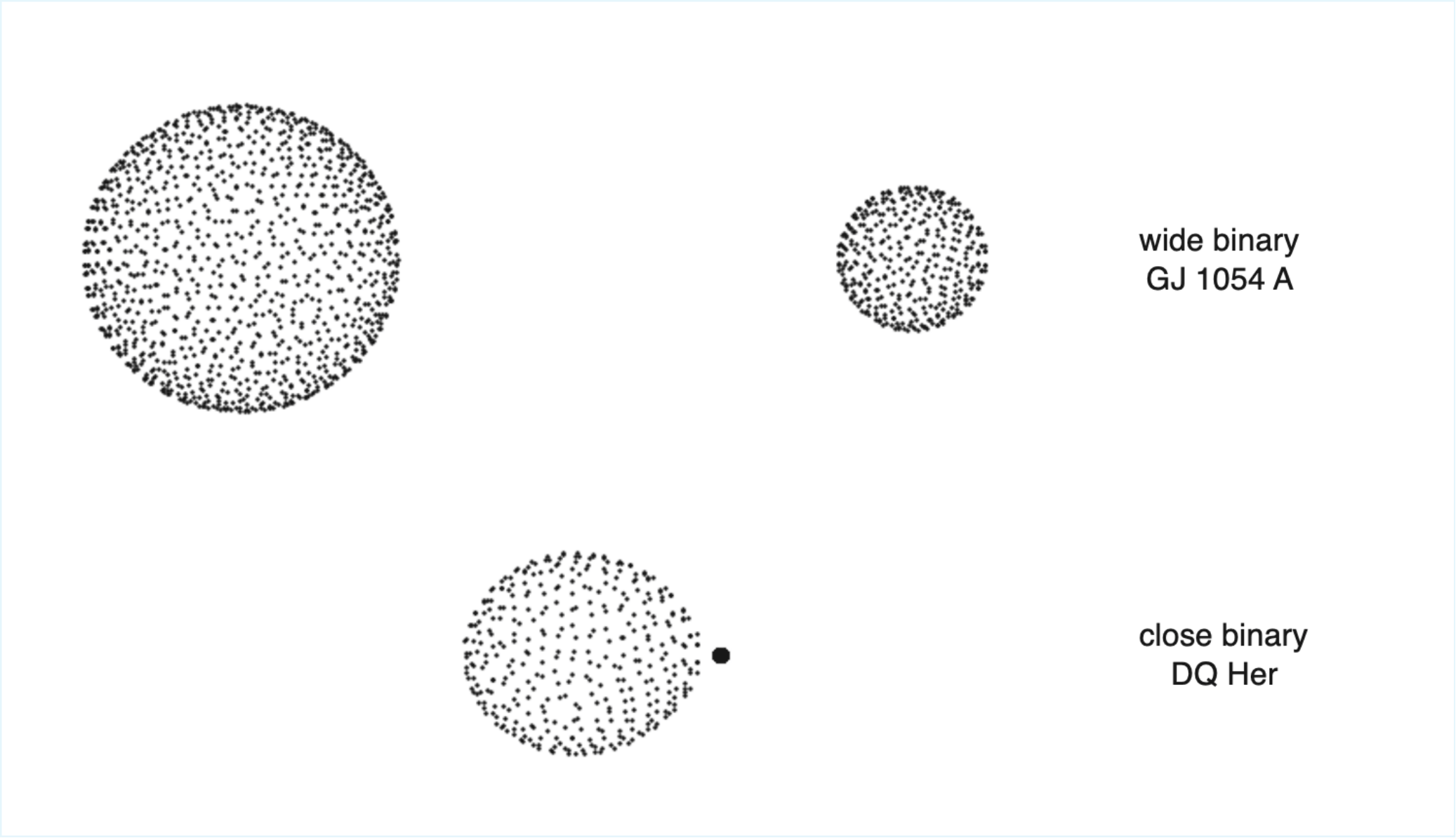}
    \caption{{Examples of mesh plots for two binary systems.} The wide binary GJ 1054 A consists of two near-spherical components, while the close binary DQ Her contains a strongly deformed M star. The stellar radii and binary separations are shown to scale.} 
    \label{figureS1.fig}
\end{figure*}



\end{appendix}

\end{multicols}
\end{document}